\documentclass[12pt]{iopart}

\usepackage{iopams}  

\usepackage[english]{babel}
\usepackage[utf8]{inputenc}
\usepackage{graphicx}
\usepackage{braket}

\expandafter\let\csname equation*\endcsname\relax
\expandafter\let\csname endequation*\endcsname\relax
\usepackage{amsmath}

\usepackage{bm}
\usepackage{xcolor}
\usepackage{enumerate}
\usepackage{multirow}
\usepackage{booktabs}

\begin{document}
\title[Glass--like transition described by toppling of stability hierarchy]{Glass--like transition described by toppling of stability hierarchy}\footnote{Dedicated to the memory of Fritz Haake}

\author{Jacek Grela$^1$ 
and Boris A Khoruzhenko $^2$}

\address{$^1$ {Institute of Theoretical Physics, Jagiellonian University, 30-348 Krakow, Poland}}
\address{$^2$ {School of Mathematical Sciences, Queen Mary University of London, London E1 4NS, UK}}
\ead{jacekgrela@gmail.com, b.khoruzhenko@qmul.ac.uk}

\begin{abstract}
Building on the work of Fyodorov (2004) and Fyodorov and Nadal (2012) we examine the critical behaviour of population of saddles with fixed instability index $k$ in high dimensional random energy landscapes. Such landscapes consist of a parabolic confining potential and a random part in $N\gg 1$ dimensions. When the relative strength $m$ of the parabolic part is decreasing below a critical value $m_c$, the random energy landscapes exhibit a glass-like transition from a simple phase with very few critical points to a complex phase with the energy surface having exponentially many critical points.  We obtain the annealed probability distribution of the instability index $k$ by working out the mean size of the population of saddles with index $k$ relative to the mean size of the entire population of critical points and observe  toppling of stability hierarchy which accompanies the underlying glass-like transition. In the transition region $m=m_c + \delta N^{-1/2}$ the typical instability index scales as $k = \kappa N^{1/4}$ and the toppling mechanism affects whole instability index distribution, in particular the most probable value of $\kappa$ changes from $\kappa = 0$ in the simple phase ($\delta > 0 $) to a non-zero value $ \kappa_{\max} \propto (-\delta)^{3/2}$ in the complex phase ($\delta < 0$). We also show that a similar phenomenon  is observed in random landscapes with an additional fixed energy constraint and in the $p$-spin spherical model. 
\end{abstract}


\maketitle

%

\section{Introduction}
\label{intro}
Low-dimensional random fields for a long time held a prominent role in sciences as many natural phenomena are described through statistics of random fields defined on either space or space--time \cite{CHRISTAKOS} with typical dimensionalities $N=2,3$ or $4$. On the other hand, only recently high-dimensional spaces ($N\gg 1$) received attention with the advent of machine learning \cite{GBCBOOK} or in studies of complex systems like spin glasses and large bio-molecules \cite{WALES}. In such applications, the dimensionality $N$ is identified with the number of degrees of freedom and the field itself is interpreted as either energy function (spin glasses, proteins) or loss function (machine learning). As was pointed out and utilized numerous times \cite{BALLARD,LUCAS,CHOROMANSKA}, the task is always to find a configuration of the system such that the energy (or loss) function is minimized. 

Although this task is straightforward in low dimensions, its successful completion for non-convex surfaces in high-dimensional spaces is notoriously intractable and is a NP-hard problem. As a result, \textit{heuristic} methods, e.g., algorithms which work approximately but whose robustness remain elusive, are widespread and remain the main choice of tool for practitioners. A well-known example is the stochastic gradient descent (SGD) algorithm derived from a gradient descent method for solving convex problems. 

Almost all problems with minimizing non-convex functions are ultimately related to the structure of their surface which varies just as the Earth's landscape does with plateaus, valleys and peaks only with a greater degree of variability. On top of that, the landscape described is very high-dimensional and thus heavily impeding our intuition. Despite these difficulties, insights into behaviour of energy/loss landscapes are crucial in understanding of the success of algorithms like SGD. In this context, the most natural approach is to study quantities related to stationary points of the function by either counting their total number or study their respective positions.  

In recent years, perhaps the most startling case of a successful approach to non-convex optimization is that of Deep Neural Networks (DNNs). Not only the success of training DNNs through SGD is surprising, also such models do not suffer from \textit{overfitting} despite being extremely overparametrized, generalize well to unseen examples and do not get stuck in local minima \cite{LECUNBOOK}. One possible explanation of these features \cite{CHOROMANSKA} is provided by inspecting statistics of stationary points of the DNN loss function based on an explicit link to spherical spin glass models. This approach along with related works \cite{AUFFINGER1,AUFFINGER2} offers a possible explanation of trainability aspects of DNNs through analyzing structure of stationary points. Our work extends this line of inquiry through a probabilistic approach to the description of populations of stationary points having a fixed instability index.

The aim of our work is to offer a detailed analysis of the stationary points 
in the energy landscapes of systems undergoing glass-like transition. The existing body of work on this subject mostly focuses on parameter ranges far from the critical threshold, both in the topologically non-trivial phase (where stationary points are exponentially abundant) and in the topologically trivial phase (where, typically, there are very few stationary points) \cite{FYOD1,BRAYDEAN1,FYODWILL1,FYOD2}.  
To the best of our knowledge, the only work providing insights into what happens near the critical threshold is one by Fyodorov and Nadal \cite{FYODNADAL1} who counted minima. We complement this study by tracking the relative sizes of populations of minima, maxima as well as saddle points with any given instability index $k$ (number of unstable directions)  {\it near the critical threshold}. To this end, we work out fractional probabilities $p_k$ for a stationary point to have a given instability index $k$, an approach introduced in \cite{KFBA2}. These probabilities are quotients of the population size of the stationary points with index $k$ to the total number of the stationary points. The picture that is emerging from our analysis offers an alternative explanation of the glass-like transition based on \emph{toppling of stability hierarchy} of populations of stationary points. In this context, the stability hierarchy refers to a steady decrease of $p_k$ when the  instability index $k$ is increasing, so that local minima are the most likely stationary points, while toppling refers to a sudden change when the most likely stationary points are those with a fixed number of unstable directions $k_{max}>0$.  
 As we show in the current work, this phenomenon which marks a global change in the underlying random landscape is shared by random landscapes, its energy constrained variant and by the paradigmatic spherical $p$-spin model.

\section{Main results}
\label{main}
In order to gain insights into statistics of stationary points in the transition region, we employ the paradigmatic model \cite{FRANZMEZARD, FYOD1} of random energy landscape
where a random scalar field is coupled to a parabolic confining potential with the transition driven by the coupling strength $\mu >0$:
\begin{equation}\label{potential}
    E(\mathbf{x}) = \frac{\mu}{2} |\mathbf{x}|^2 +V(\mathbf{x})\, .
\end{equation}
Here $\textbf{x}$ is a vector in $N$-dimensional state-space and $V(\textbf{x})$ is isotropic homogeneous Gaussian vector field with zero mean value and covariance function 
\begin{align}
\label{covariance}
\langle V(\textbf{x}) V(\textbf{x}') \rangle = N f\left (\frac{|\textbf{x}-\textbf{x}'|^2 }{2N} \right ).
\end{align} 

The main feature of model \eqref{potential} is the existence of two distinct phases in the thermodynamic limit $N \to \infty$ with a sharp transition region between the two phases at $\mu_c = \sqrt{f''(0)}$ \cite{FYOD1}. Introducing the rescaled coupling strength 
\begin{align}\label{m}
m = \frac{\mu}{\sqrt{f''(0)}}\, ,
\end{align} 
for large values of $m$ the system is in a topologically trivial phase whereby the parabolic confining potential  
dominates the energy landscape (the probability of finding more than one stationary point is zero in the limit $N\to \infty$). 
As $m$ decreases below the critical threshold $m_c=1$ the energy landscape becomes highly complex which 
manifests itself in an exponential explosion of stationary points - the (average) total number of stationary points and the number of local minima grow exponentially as the dimension of the state space $N$ increases. 
This transition 
is frequently called glass--like as it is closely related to one found in spherical spin--glasses \cite{BENAROUS}. 

In a natural way, the energy landscape \eqref{potential} gives rise to a gradient flow which is defined by the differential equation 
\begin{equation}\label{b1}
\dot{\mathbf{x}} = - \nabla E(\mathbf{x})\, .
\end{equation} 
Then the stationary points $\mathbf{x}_*$ of $E(\mathbf{x})$, i.e., the points where $\nabla E$ vanishes, are the equilibria (fixed points) of the gradient flow \eqref{b1}. In this picture, if $\mathbf{x}_*$ is a point of local minimum of $E(\mathbf{x})$ (i.e., all eigenvalues of the Hessian $(\partial_i\partial_j E(\mathbf{x}))_{ij}$ at $\mathbf{x}=\mathbf{x}_*$ are positive)  then the equilibrium $\mathbf{x}_*$  is asymptotically stable. That is, a small displacement from $\mathbf{x}_*$ in any direction results in the system asymptotically returning back to $\mathbf{x}_*$. If $\mathbf{x}_*$  is a saddle and $k$ is the number of negative eigenvalues of the Hessian of $E(\mathbf{x})$ at $\mathbf{x}=\mathbf{x}_*$ then the equilibrium at $\mathbf{x}_*$ will have $N-k$ stable directions. Displacement along these directions will result in the system asymptotically returning back. In this way, the index $k$ is a measure of instability of the equilibrium, and we shall call it the instability index. The higher the value of $k$ is, the fewer stable directions there will be at the equilibrium. Obviously, local maxima, i.e., stationary points where $k=N$, are most unstable.

\subsection{Populations of stationary points}
\label{Cohorts}

In this work, we go beyond counting few sub--populations (like minima) of stationary points in absolute terms and instead work out relative or fractional probability distributions of  stationary points with fixed instability index. This approach enables novel, refined questions like:
\begin{itemize}
\item Given a randomly sampled stationary point, what would be its most likely instability index? How does a full probability distribution of indices look like?
\item How does the probability distribution of indices change in the vicinity of transition region?
\end{itemize}

Such pertinent enquiries can be made by a local agent (like an SGD algorithm or glassy system looking for a configuration minimizing its energy) probing the landscape and encountering saddles on its way. 
For example, it was argued in ref.~\cite{Kurchan_1996} that the abundance of saddles with a large number of stable directions leads to slowing down the gradient descent dynamics in high dimensional energy landscapes due to the dominance of borders in high dimensions: at low temperatures the system is trapped for long times near borders (ridges) of basins of attraction of local minima and the gradient descent is determined mainly by nearby saddles.

In what follows, we utilize three interrelated counting statistics: 
\begin{itemize}
\item $\mathcal{N}_k$, the number of stationary 
points with instability index $k$;
\item $\mathcal{N}_{\text{eq}} = \sum_{k=0}^N \mathcal{N}_k$, the total number of stationary points;

\item $\mathcal{N}^{(k)} = \sum_{n=0}^{k} \mathcal{N}_n $ the number of stationary points with instability index up to $k$.
\end{itemize}
The quotient of the first two counting statistics is the relative frequency $\mathcal{N}_k /\mathcal{N}_{\text{eq}}$ of saddles with instability index $k$, whilst the last one is the associated cumulative frequency distribution $\mathcal{N}^{(k)}/\mathcal{N}_{\text{eq}}$. 
If $\textbf{x}_*$ is a stationary point of $E(\mathbf{x})$ drawn at random from the entire population of stationary points then, as was shown in \cite{KFBA2},  the probability for $\textbf{x}_*$ to have $k$ unstable directions is given by the average of $\mathcal{N}_k /\mathcal{N}_{\text{eq}}$  over the realizations of the random field $V(\textbf{x})$:
\begin{equation}\label{p_k_quenched}
\Pr  \{\text{$\textbf{x}_*$  has instability index $k$}\} =\langle { \mathcal{N}_k}/{\mathcal{N}_{\text{eq}}} \rangle \, .
\end{equation}
Therefore, $\left < \mathcal{N}_k/\mathcal{N}_{\text{eq}} \right >$ and $\left < \mathcal{N}^{(k)}/\mathcal{N}_{\text{eq}} \right >$ are, respectively, the probability density function (pdf) and cumulative distribution function of the instability index $k$.  

In the context of the above two questions, calculating both averages is a natural starting point. This seems a prohibitively difficult task, and, instead, we set out to  
analyse the \textit{annealed} probabilities 
\begin{equation}
p_k  = \frac{\langle \mathcal{N}_k \rangle}{\langle \mathcal{N}_{\text{eq}} \rangle}, \quad
P_{k} = \frac{\langle \mathcal{N}^{(k)}\rangle }{\langle \mathcal{N}_{\text{eq}}\rangle},  \label{pk}
\end{equation}
where enumerator and denominator are averaged separately and combined afterwards. To justify a connection between the right-hand side in \eqref{p_k_quenched} and its annealed counterpart $p_k$, counting statistics in both enumerator and denominator ought to have a self-averaging property in the limit of high dimensionality limit. The recent works \cite{SUBAG,RBC1,Auf2020b} addressed this type of question for the pure $p$-spin spherical model whose energy landscape falls in the same class as the random energy model and gave an affirmative answer for equilibria with sufficiently low energy or with finite instability index $k$. This gives rise to a hope that for some classes of coupling fields the annealed picture will resemble the quenched one. While the task of identifying such classes of coupling fields is a challenging open problem which deserves further investigations, we think that the annealed probabilities deserve a closer look. The picture which is emerging from our analysis of these probabilities exhibits some interesting features and is described below.

\begin{figure}
\includegraphics[scale=.62]{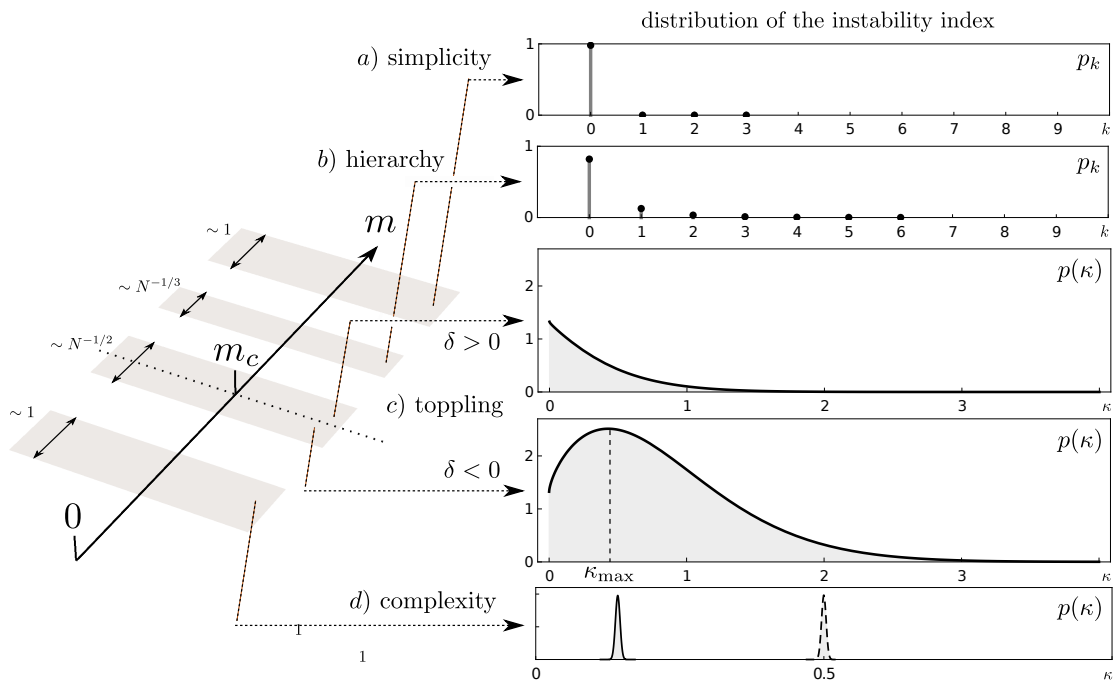} 
\caption{
Diagram illustrating   
the toppling mechanism in 
the random energy landscape model (\ref{potential}) -- (\ref{covariance}) in the limit of high dimensionality $N\gg 1$. 
The diagram on the left-hand side depicts the four scaling regions of parameter $m$ (\ref{m}).
The vertical array of five plots on the right-hand side depicts the annealed distribution of the instability index $k$ in each of these regions (there are two plots for the toppling region).
The plot at the top depicts the annealed probabilities $p_k$ (\ref{pk})  in the simplicity region and the plot below it depicts $p_k$ in the hierarchy region. In these two regions the typical values of $k$ are finite and, correspondingly,  the plots are discrete. 
%
In the toppling 
and complexity regions the typical values of $k$ scale with, correspondingly, $N^{1/4}$ and $N$. 
There, the annealed distribution of $k$ is best described via the cumulative probability $P_k=\sum_{n=0}^k p_n$. The corresponding densities, $ p^{(c)}\! (\kappa)=\frac{d}{d\kappa} P_{\kappa N^{1/4}}$  and $ p^{(d)}\! (\kappa)=\frac{d}{d\kappa} P_{\kappa N}$, are depicted in the bottom three plots. 
In the toppling region 
the annealed density $p^{(c)}\!(\kappa)$ undergoes a gradual change when the value of $m$ decreases below the critical threshold $m_c=1$ from being a monotone decreasing function of $\kappa$ (the likeliest saddles are local minima) to being a unimodal function  (the likeliest saddles have $\kappa_{\max} N^{1/4}$ unstable directions, $\kappa_{\max} = \frac{4\sqrt{2}}{3\pi} (- \delta)^{3/2}$). 
The plots were produced using analytic expressions presented in the third column in Table~\ref{tab3} and evaluated at parameter values $m>1$ in plot a), $\delta=0.8$ in plot b), $\delta=0.2$ and $\delta=-0.8$ in  plot c), and $m=0.6$ (solid line) and $m=0$ (dashed line) in plot d). 
Plot b) was produced with the help of  numerical package \cite{BORNEMANN}. 
}
\label{fig1}
\end{figure}

\subsection{Toppling of stability hierarchy}
\label{evolution}

As was mentioned above, one manifestation of the phase transition in the random energy model \eqref{potential}--\eqref{covariance}  is the exponential explosion in the number of stationary points. In the limit of high dimensionality
$N\to\infty$, the complexity exponents 
\begin{align}\label{exponents}
\Sigma_{\text{eq}} = \lim_{N\to\infty}\frac{1}{N} \ln\, \langle \mathcal{N}_{\text{eq}} \rangle \quad  \mathrm{ and } \quad \Sigma_{0} =\lim_{N\to\infty} \frac{1}{N} \log \langle \mathcal{N}_{0} \rangle
\end{align}
associated, respectively, with the total number of stationary points and the number of local minima, are positive for every $m<1$,
\begin{align}\label{Sigma_eq}
\Sigma_{\text{eq}} = \frac{m^2-1}{2}-\ln m, \quad \Sigma_{0}=\Sigma_{\text{eq}} -(1-m)^2\, ,\qquad (0<m<1)
\end{align}
and vansih for every $m>1$ \cite{FYOD1,FYODNADAL1}. However, as can be seen from  \eqref{Sigma_eq} at the critical threshold these two counting statistics develop the exponential growth on different scales: the width of the transition region for $\left < \mathcal{N}_{\text{eq}} \right >$ is $N^{-1/2}$, whilst the width of the transition region for $\left < \mathcal{N}_{0} \right >$ is  $N^{-1/3}$. When one analyzes the relative sizes of populations of stationary points near the critical threshold, such as the annealed probabilities $p_k$ and $P_k$ \eqref{pk}, the microscopic scales get superimposed. This suggests that the glassy transition in model \eqref{potential}--\eqref{covariance}  has several distinct transition regions which we will now describe.

It is convenient to encode the scaling regimes by the formula 
 \begin{align*}
 m= 1 + \frac{\delta}{N^\beta}.
 \end{align*}
 The parameter values $\beta=1/2$ and $\beta=1/3$ define the two aforementioned microscopic scales  and the parameter value $\beta=0$ defines a global scale. Based on these three natural scales, we can identify four distinct scaling regions of change consisting of two microscopic and two macroscopic scales, see Figure \ref{fig1} and Table~\ref{tab3}:

\begin{enumerate}[a)]
\item \textit{Simplicity region}, $m= 1+ \delta>0$ \\[0.5ex]
In this region 
the parabolic confining potential dominates in the limit $N\to\infty$. The instability index $k$ is a discrete variable and $k=0$ corresponds to a minima. With probability asymptically close to 1 the system has only one stationary point which is a local minimum, and so $p_k = 1$ in the limit $N\to\infty$ if $k=0$ and $p_k = 0$ otherwise. This is illustrated 
in plot a), Figure \ref{fig1}.  

\item \textit{Hierarchy region}, $m= 1+\delta /N^{1/3}$, $\delta> 0$\\[0.5ex]
As the value of $m$ is decreasing and getting closer to the critical threshold 
$m_c=1$, 
the system exits the simplicity region and enters a hierarchy region. In this region the mean number of saddles $\langle \mathcal{N}_{\text{eq}} \rangle$ is increasing as $m$ gets closer to $m_c$ (but staying finite in the limit of high dimensionality) and the deviations of the instability index $k$ from zero become larger. This 
results in a flow of indices away from $k=0$ 
and the emergence of groups of increasingly more unstable saddles as evidenced by non-zero values of $\langle \mathcal{N}_k \rangle$ for $k>0$, see Table~\ref{tab3}. 
Consequently, the annealed probability distribution of the instability index $k$ develops a non-zero tail extending to finite values of $k$ and displaying the hierarchy of stability $p_0>p_1>p_2 > \ldots $ with the most likely stationary point being a local minimum. This is illustrated in plot b),  Figure \ref{fig1}.  Note that typical values of the instability index $k$ do not scale with $N$ in the limit of high dimensionality,  i.e. $k$ takes finite values $k=0,1,2, \dots $. 

\item \textit{Toppling region}, $m =1+  \delta /N^{1/2}$, $\delta \in \mathbb{R}$ \\[0.5ex]
As the value of $m$ is decreasing further and getting closer to the critical threshold, the flow of indices away from zero becomes stronger and their distribution flatter, i.e., the difference between $p_k$ and $p_{k+1}$ is becoming smaller and smaller. This microscopic mechanism is fundamental 
and eventually leads to a macroscopic change which occurs in the region $m =1+  \delta /N^{1/2}$. It can be shown that in this region the total number of stationary points scales as $N^{1/4}$ and so do the typical values of the instability index $k$. It is then natural to introduce rescaled instability index  $\kappa =k/N^{1/4}$ which becomes a continuous random variable in the limit of high dimensionality.   It is instructive to inspect the dependence of the rescaled  index density  
 $\frac{d}{d\kappa} P_{\kappa N^{1/4}}$ 
 on parameter $\delta $, see Figure~\ref{fig1} and Table~\ref{tab3}. For every fixed $\delta >0$, this density 
 is monotonically decreasing function of $\kappa$ in the interval $0\le \kappa < \infty$ and, hence, the hierarchy developed in region b) persists.   The critical threshold $m_c=1$ which corresponds to  $\delta=0$ on this microscopic scale is the tipping point. For every fixed $\delta<0$, the index density is a unimodal function of $\kappa$ attaining its maximum value at $\kappa_{\max} = \frac{4\sqrt{2}}{3\pi} (- \delta)^{3/2}$.
 The loss of monotonicity 
 means that the most likely stationary point is no longer a local minimum but instead a saddle with $\kappa_{max}N^{1/4}$ unstable directions. In other words, the hierarchy of stationary point populations is broken and transition to complex phase starts which eventually produces a typical (i.e. most probable) stationary point with non-zero instability index.

\item \textit{Complexity region}, $0< m <1$  \\[0.5ex]
As the scaled coupling strength $m$ is decreasing further below the critical threshold, the system enters the topologically non-trivial phase where the total number of stationary points is exponential in $N$, see \eqref{Sigma_eq}.  In this region, the typical values of $k$ scale with $N$ 
and thus $k=\kappa N$. The resulting index density $\frac{d}{d\kappa} P_{\kappa N}$  has a highly localized 
peak at 
\begin{align*}
\kappa_{max}(m)&=\int_m^1 \rho_{\text{sc}} (\lambda) d\lambda\, , \quad\quad\quad \rho_{\text{sc}} (\lambda)= \frac{2}{\pi} \sqrt{1-\lambda^2}  \\
& = \frac{1}{\pi} \Big( \arccos m - m\sqrt{1-m^2} \Big)
\end{align*}
see plot d), Figure~\ref{fig1}. As one would expect 
$\kappa_{max}(1)=0$ and $\kappa_{max}(0)=1/2$, so that 
when the parabolic confining potential is turned off completely, the most likely instability index is $N/2$. This of course makes complete sense as for a pure random field the stable and unstable directions are equally likely. 

\end{enumerate}

Closed form expressions for the fractional probabilities \eqref{pk} in the limit of high dimensionality $N\to\infty$ are written down explicitly in Table \ref{tab3} while all the derivations can be found in \ref{app0}.

\begin{table}
\caption{\label{tab3} Annealed cumulative distribution $\displaystyle{P_k={ \langle \mathcal{N}^{(k)} \rangle}/{ \langle \mathcal{N}_{\text{eq}} \rangle}}$  of the instability index $k$ in the random energy landscape model \eqref{potential}--\eqref{covariance}
in the limit of high dimensionality $N\gg 1$. 
Here,  $\rho_{\text{edge}}(\lambda)=\left ( \text{Ai}'(\lambda) \right )^2 - \lambda (\text{Ai}(\lambda))^2 + \frac{1}{2} \text{Ai}(\lambda) \left (1 - \int_\lambda^\infty \text{Ai}(t) dt \right )$ with $\text{Ai}$ denoting the Airy function, and $\rho_{\text{sc}}(\lambda)=\frac{2}{\pi}\sqrt{1-\lambda^2}$  are the eigenvalue densities at the edge and in the bulk of the eigenvalue distribution in the GOE, and $F_n(\lambda)$ is the cdf of the top $(n+1)$-st eigenvalue in the GOE. The complexity exponent $\Sigma_{\text{eq}}$ is given in ~\eqref{Sigma_eq} and $c=\big({3\pi}/{(4\sqrt{2})}\big)^{2/3}$. The probability densities shown in the third column are plotted in Figure \ref{fig1}.
}
\footnotesize\rm
\begin{tabular}{@{}lllll}
\br
Coupling      & Cumulative & Density of  & Total number of  \\[0.1ex]
 strength   $m$    & distribution $P_k$&  distribution &  saddles  $\langle \mathcal{N}_{\text{eq}} \rangle$ \\[0.1ex]
\mr
$m>1$ 
&  
$\sum_{n=0}^{k} p_n$ 
& $p_n=\delta_{n,0}$ & 1\\
\addlinespace[3ex]
$
\displaystyle{
m =  1+ \frac{\delta}{\sqrt[3]{N} }, \,
\delta>0
}
$  
& $\sum_{n=0}^{k} p_n$ &
$\displaystyle{
p_n=
\frac
{\int_{-\infty}^\infty e^{\delta \lambda} dF_n(\lambda) }
{\int_{-\infty}^\infty e^{\delta\lambda} \rho_{\text{edge}}(\lambda)d\lambda} 
}$ & $\displaystyle{
2e^{-\frac{\delta^3}{3}}\!\!\int_{-\infty}^\infty e^{\delta\lambda} \rho_{\text{edge}}(\lambda)d\lambda
}$\\
\addlinespace[3ex]
$
\displaystyle{
m  =  1+  \frac{\delta}{\sqrt{N} },\, 
\delta \in \mathbb{R}
}
$
& 
$\displaystyle{
\int_0^{\frac{k}{\sqrt[4]{N}}} \!p(x) dx 
}$
&
$\displaystyle{
p(x) \!=\! c\frac
{e^{-\big(\delta+c x^{{2}/\!{3}}\big)^{\!2}}}
{\int_0^{+\infty}e^{-\big(\delta+c x^{{2}/\!{3}}\big)^{\!2}}\!\!dx }
 }$ & $\displaystyle{
2N^{1\!/4}e^{\delta^2}\!\!\!\int_0^{+\infty}\!\!\!e^{-\big(\delta+cx^{{2}/\!{3}}\big)^{\!2}}\!\!\!dx
}$\\
 \addlinespace[3ex]
$0<m<1$ & 
$\displaystyle{
\int_0^{k\!/\!N}\!\!\!p(x) dx 
}$
&
$\displaystyle{
p(x) \!= \! \delta \Big(x \!-\! \int_{m}^1 \rho_{\text{sc}}(\lambda) d\lambda\Big)
}$ &$\displaystyle{
\sqrt{4\pi N} \rho_{\text{sc}} (m) e^{N\Sigma_{\text{eq}}(m)} 
}$\\
\br
\end{tabular}
\end{table}

\subsection{Universality of the toppling mechanism}

Although in this work we describe the glass-like transition via the toppling of stability hierarchy for the toy model \eqref{potential}, we believe this mechanism holds more generally. To support this claim, in  \ref{fixeden} we report on an analogous mechanisms driving the phase transition in the model \eqref{potential} with an additional fixed energy constraint \cite{FYODWILL1} and in and \ref{spherical}  we do the same for the $p$-spin spherical model \cite{KTJSPIN,CSSPIN,FYOD2,AUFFINGER1}. Below we provide a brief summary of our findings. 

\begin{figure}
\includegraphics[scale=.62]{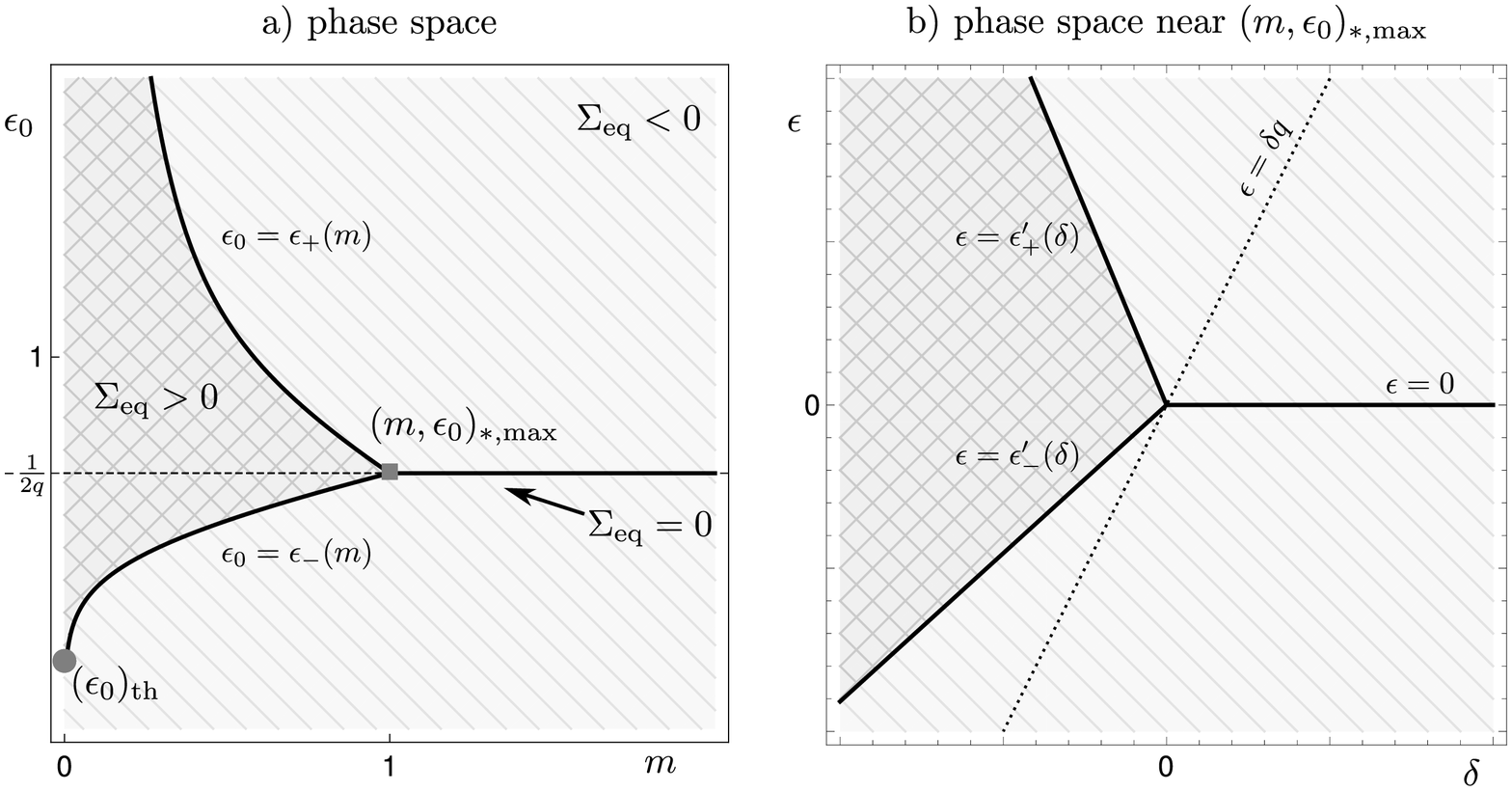}
\caption{Phase space in the fixed energy model \eqref{potential} -- \eqref{covariance}. The macroscopic phase diagram in the $(m,\epsilon_0)$-plane is depicted in the plot on the left. The point $(m,\epsilon_0)_{*,\max}=(1,-\frac{1}{2q})$, represented by the grey square on the plot, is the critical point. The criss-crossed region to the left of this point which is bounded by the lines \eqref{curves} is the complexity region. The grey dot marks the threshold energy $(\epsilon_0)_{\text{th}} = \epsilon_{-}(0)= - \frac{1+2q^2}{2q}$ below which the random energy landscape has no stationary points for any $m>0$. 
The macroscopic toppling region around the critical point, defined in \eqref{tr}, is depicted in the plot on the right. The dotted line is the boundary line along which toppling of the hierarchy of stability happens, see items a), (b), c) and d) above and Figure~\ref{fig1}}
\label{fig2a}
\end{figure}

First, consider the same random energy landscape model \eqref{potential} -- \eqref{covariance} as before but now at a fixed energy level $E_0$,
\begin{align}\label{E0}
 E(\textbf{x}) = E_0 \, .
 \end{align}
We shall call this model the fixed energy model or the constrained random landscape model. It has two parameters. One is the coupling strength $\mu$ and the other one is the energy level $E_0$. Similarly to the unconstrained model, one can investigate the complexity exponent $\Sigma_{\text{eq}} $ associated with the total number of stationary points $n_{\text{eq}}$ at energy level $E_0$,
$
\Sigma_{\text{eq}} = \lim_{N\to\infty}\frac{1}{N} \ln\, \langle n_{\text{eq}} \rangle 
$,
and, more generally, the relative (to the total number) average sizes of the population of saddles at energy level $E_0$ with instability index $k$. The latter is described by the annealed cumulative distribution function 
$
P_k = {\langle n^{(k)} \rangle}/{\langle n_{\text{eq}} \rangle}
$
of the instability index $k$,
where $n^{(k)} $ is the number of stationary points at energy level $E_0$ with instability index up to $k$. 

Instead of extensive parameters $\mu$ and $E_0$ it is more convenient to use their intensive versions, 
 the rescaled coupling strength $m$ (\ref{m}) and the rescaled energy level 
\begin{align}\label{epsilon0}
 \epsilon_0 = \frac{E_0}{ N\sqrt{f(0)}}\, . 
  \end{align}
The complexity exponent $\Sigma_{\text{eq}} $ can be obtained in a closed form. Referring the reader to  \ref{fixeden} for details, here we focus on the macroscopic phase diagram which emerges from this calculation, see the left plot in Figure \ref{fig2a}. The macroscopic phase space is defined by the zero level line of the complexity exponent $\Sigma_{\text{eq}}$ in the $(m,\epsilon_0)$-plane. This line consists of two curves (defined in \eqref{curvesA})
\begin{align}\label{curves}
\varepsilon_0=\varepsilon_{\pm} (m), \quad 0\le m \le 1\, ,
\end{align}
and a straight line 
\begin{align}\label{line}
\varepsilon_0=-\frac{1}{2q}, \quad m \ge  1,
\end{align}
where 
\begin{align*}
q = -\frac{\sqrt{f^{\prime\prime} (0) f(0)}}{ f^{\prime} (0)}>0\, . 
\end{align*}
All three elements intersect at a point  $(m,\epsilon_0)_{*,\text{max}}=(1,-\frac{1}{2q})$. The complexity region in the fixed energy model  is the area in the $(m,\epsilon_0)$-plane which is bounded by the straight line $m=0$ and the two curves \eqref{curves}. For all parameter values inside this region the complexity exponent $\Sigma_{\text{eq}} $ is positive and the random energy landscape  \eqref{potential} -- \eqref{covariance} has, on average, exponentially many stationary points at energy levels in the interval $\epsilon_{-}(m)< \epsilon_0 < \epsilon_{+}(m)$. 
With the exception of the straight line \eqref{line}, the complexity exponent $\Sigma_{\text{eq}} $ is negative outside the complexity region. This is the two-dimensional simplicity region. For all parameter values $(m,\epsilon_0)$ in this region the probability for the random energy landscape to have at least one stationary point is exponentially small unless  $\epsilon_0=-\frac{1}{2q}$ in which case the landscape typically has one stationary point (minimum) at each $m>1$. 
 
The glass-like  transition from the simplicity to complexity phases happens at the critical point $(m,\epsilon_0)_{*,\text{max}}$. In what follows we focus only on the toppling region, depicted in the right plot in Figure \ref{fig2a}, which is an area of linear size $O(\frac{1}{\sqrt{N}} )$ around this point:
\begin{align}\label{tr}
m=1+\frac{\delta}{\sqrt{N}}, \quad \epsilon_0=-\frac{1}{2q}+\frac{\epsilon}{\sqrt{N}}, \quad\quad \delta,\epsilon =O(1). 
\end{align}
The toppling mechanism manifests itself in the form of annealed cumulative distribution of instability index:
\begin{align}
\label{Pkconstr}
    P_{k} (\delta,\epsilon) \sim  
    \frac
{ \int_0^{\frac{k}{\sqrt[4]{N}}}  e^{- \big(\Delta_q(\delta, \epsilon) +c_q x^{{2}/\!{3}}\big)^{\!2}}  dx}
{\int_0^{+\infty}e^{-\big(\Delta_q(\delta, \epsilon)+c_q x^{{2}/\!{3}}\big)^{\!2}}\!\!dx } \, ,
  \end{align}
where $ c_q = \sqrt{\frac{2q^2+3}{2q^2-1}} \left ( \frac{3\pi}{4\sqrt{2}} \right )^{2/3}$ is a constant and 
 \begin{align*}
     \Delta_q(\delta, \epsilon) = \frac{2\sqrt{2}q^2 }{\sqrt{(2q^2-1)(q^2+2)}}\left(\delta - \frac{\epsilon}{q}\right).
 \end{align*}
The behaviour of function \eqref{Pkconstr} is driven by the sign of $\Delta_q(\delta, \epsilon)$ so the condition $\Delta_q(\delta, \epsilon)=0$ describes a boundary line along which toppling takes place (the dotted line in right plot of Figure \ref{fig2a}). In the $(\delta, \epsilon)$-plane the transition from the simplicity to the complexity regions will occur along any path  that starts to the right of the boundary line $\epsilon = q\delta$ in the region $\delta \gg 1 $, enters the cone bounded by two straight lines  
\begin{align*}
\epsilon^{\prime}_\pm (\delta) = \delta \left ( - \frac{3}{2q} \pm \frac{\sqrt{3+2q^2}}{q} \right ),\, \quad \quad \delta \le 0, 
\end{align*}
and then continues inside this cone to the region of  $\delta\ll -1$. 
When we take into account only such meaningful paths, we recreate the toppling mechanism in two-dimensional phase space. In particular, the system develops the most probable non-zero instability index $\kappa'_{\max} = \Big(\!\!-\frac{\Delta_q(\delta, \epsilon)}{2c_q}\Big)^{3/2}$ on crossing the line $\epsilon=q\delta$ (dotted line in the right plot in Figure \ref{fig2a}) while the main features of the annealed probability density of the instability index mirrors that of the unconstrained model presented in Figure \ref{fig1}. Furthermore, connection with the unconstrained model is evident as the annealed probability density in the toppling region in this model  model have the same functional form as  \eqref{Pkconstr} (compare with the relevant entries of Table \ref{tab3}).

 \bigskip

\begin{figure}
\includegraphics[scale=.62]{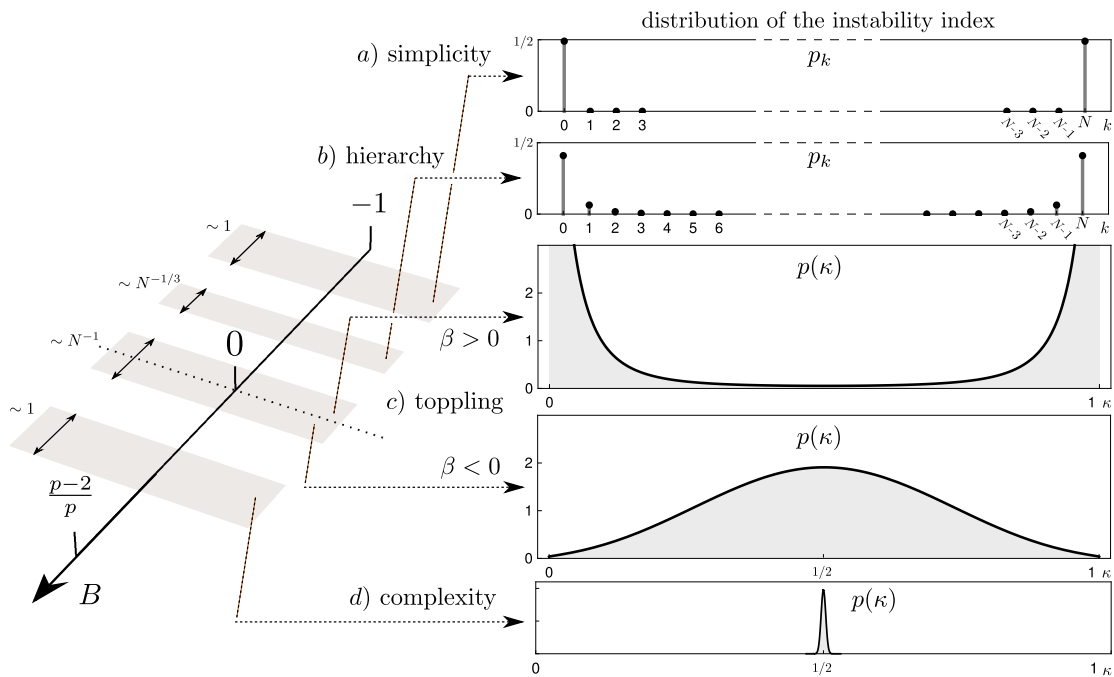} 
\caption{ 
Diagram illustrating  
toppling mechanism in 
the $p$-spin spherical model (\ref{spinglass}) -- (\ref{spinglass1})  in the limit of high dimensionality $N\gg 1$. 
The diagram on the left-hand side depicts the four scaling regions of parameter $B$ (\ref{B}).
The vertical array of five plots on the right-hand side depicts the annealed distribution of the instability index $k$ in each of these regions (there are two plots for the toppling region).
The top two plots depicts the annealed probabilities $p_k$ (\ref{pk})  in the simplicity and hierarchy regions where the typical values of $k$ are finite.
In the toppling 
and complexity regions the typical values of $k$ scale with $N$.
 There, the annealed distribution of $k$ is best described via the cumulative probability $P_k=\sum_{n=0}^n p_n$. The corresponding density $p(\kappa)= \frac{d}{d\kappa} P_{\kappa N}$ is depicted in the bottom three plots.  In the toppling region 
 $p(\kappa)$ undergoes a gradual change when the parameter $B$ increases above the critical threshold $B=0$ from being a monotone decreasing function on the interval $[0,1/2]$  (the likeliest saddles are local minima and maxima) to being a monotone increasing function (the likeliest saddles have $N/2$ unstable directions). 
The plots were produced using analytic expressions presented in the third column in Table~\ref{tab7}  and evaluated at specific parameter values 
($B = -1$ in (a), $\beta =-0.8$ in (b), $\beta=6$ in (c) and $\beta=-4 $ in (d) and $B=\frac{p-2}{p}$ in (d)).
Plot (b) was produced with the help of  numerical package \cite{BORNEMANN}. 
}
\label{fig3}
\end{figure}

Now, consider the $p$-spin spherical model defined by an energy function
\begin{align}
\label{spinglass}
E_{\circ}(\textbf{x}) = \sum_{i_1,...,i_p=1}^{N+1} J_{i_1,i_2,...,i_p} x_{i_1} x_{i_2} ... x_{i_p} + \sum_{i=1}^{N+1} h_i x_i,
\end{align}
where $\textbf{x}$ is an $N+1$ dimensional vector constrained to lie on the sphere $\sum_{i=1}^{N+1} x_i^2 = N$ and $p\geq 2$ is a positive integer. Symmetric coupling matrix $J$ and random external field $h_i$ are both drawn from Gaussian distributions with vanishing means and variances 
\begin{equation}\label{spinglass1}
 \langle(J_{i_1i_2...i_p})^2 \rangle = \frac{J^2}{pN^{p-1}}, \quad \langle h_i^2 \rangle= \sigma^2. 
\end{equation}

\begin{table} 
\caption{\label{tab7} Annealed distribution $\displaystyle{P_k={ \langle \mathcal{N}^{(k)} \rangle}/{ \langle \mathcal{N}_{\text{eq}} \rangle}}$   of the instability index $k$ in the spherical spin-glass model (\ref{spinglass})--(\ref{spinglass1}) in the limit of high dimensionality $N\gg 1$.  The first two scaling regimes are expanded for two different parameter ranges of $k$  due to topological constraint linking stability and instability in the spherical model. $\rho_{\text{edge}}(\lambda)$, $\rho_{\text{sc}}(\lambda)$ and $F_n(\lambda)$  are as defined in Table~\ref{tab3}  
and $Q_{x}$ is the quantile function of the semicircular law, $\int_{Q_{x}}^1 \rho_{\text{sc}}(\lambda) d\lambda =x $. 
The mean total number of saddles $\langle \mathcal{N}_{\text{eq}} \rangle $ in all four regimes was obtained in ref.~\cite{FYOD2}. 
}
\footnotesize\rm
\begin{tabular}{@{}lllll}
\br
 $B=\frac{J^2(p-2) - \sigma^2}{J^2 p + \sigma^2}$    & Cumulative  & Density of  distribution & Total number  of \\
    &  distribution $P_k$ & &  saddles $\langle \mathcal{N}_{\text{eq}} \rangle$ \\
\mr
$-1 < B < 0$
&  
$
P_k=\sum\limits_{n=0}^{k} p_n 
$ 
& 
$
\displaystyle{
p_n=p_{N-n}=\frac{1}{2}\, \delta_{n,0} 
}
$ & 
2\\
\addlinespace[3ex]
$
\displaystyle{
B =  - \frac{\beta}{\sqrt[3]N},  \,  \beta > 0
}
$
  & 
$
P_k=\sum\limits_{n=0}^{k} p_n 
$  &
$ 
\displaystyle{
p_n= p_{N-n}=\frac{1}{2}
\frac
{\int_{-\infty}^\infty e^{\beta \lambda} dF_n(\lambda) }
{\int_{-\infty}^\infty e^{\beta \lambda} \rho_{\text{edge}}(\lambda)d\lambda}
} 
$  
& 
$\displaystyle{4e^{-\frac{\beta^3}{3}}\!\!
\int_{-\infty}^\infty \!\!\! e^{\beta\lambda}\! \rho_{\text{edge}} (\lambda) d\lambda
}$
\\
 \addlinespace[3ex]
$
\displaystyle{
B  =  - \frac{\beta}{N}, 
\, \beta \in \mathbb{R}
}
$
& 
$
 \displaystyle{
\int_0^{{k}\!/\!{N}} \!p(x) dx 
}$
&
$\displaystyle{
p(x) \!=\! 
\frac
{e^{\, \beta Q_x^2} }
{\int_{-1}^{1} e^{\, \beta \lambda^2} \!\rho_{\text{sc}}(\lambda) d\lambda }
 }$ & 
 $
 \displaystyle{2N e^{-\beta}\!\!\int_{-1}^{1} \!\!e^{\beta \lambda^2} \!\!\!\rho_{\text{sc}} (\lambda)   d\lambda }$
\\
\addlinespace[3ex]
$\displaystyle{
0 < B \le \frac{p-2}{p}
}$  & 
$
\displaystyle{
\int_0^{k\!/\!N}\!\!p(x) dx 
}$
&
$\displaystyle{
p(x) \!=\!  \delta \Big(x- \frac{1}{2}\Big)
}$ 
&
$
\displaystyle{4\sqrt{N} \sqrt{\frac{1+B}{\pi B}} e^{\frac{N}{2}\! \log\! \frac{1+B}{1-B}}}$
\\
\br 
\end{tabular}

\end{table}

In  \ref{spherical} we both recall known results and summarize new calculations enabling calculation of asymptotic forms of annealed probabilities across regions a)-d), see Table \ref{tab7} for a summary. In Figure \ref{fig3} we plot these probabilities in all four regions as a function of an effective variable 
\begin{align}\label{B}
B = \frac{J^2(p-2) - \sigma^2}{J^2 p + \sigma^2}
\end{align} 
combining variances $J,\sigma$ and the parameter $p$. Importantly, although at first the resulting picture might not resemble Figure \ref{fig1} plotted for the toy model \eqref{potential}, it is due to development of a dual hierarchy resulting in likewise toppling of both hierarchies simultaneously.

In contrast to the toppling mechanism of model \eqref{potential} where around a single minimum in the simple region one hierarchy is developed, in the spherical model, by topological reasons, the simple phase consists instead of two stationary points -- a minimum and a maximum. These in turn produce two disjoint stability hierarchies and eventually, in the toppling region both hierarchies are toppled and merged together to create a joint density centered around scaled instability index $\kappa = 1/2$. Complexity region is trivial and centered around $\kappa=1/2$ so the toppling mechanism happens on a smaller scale. 

\section{Methods}
\label{deriv}
This section provides an outline of the approach we employ to calculate the fractional probabilities \eqref{pk} in the limit of high dimensions. Technical details of our calculations can be found in Appendices. For simplicity, we only discuss  the unconstrained random energy landscape model \eqref{potential} -- \eqref{covariance}. Our approach to the other two models, the fixed energy landscape model and the $p$-spin spherical model is similar and we only provide references to key results relevant to these two models. 


The key observation that enables calculation of the fractional probabilities \eqref{pk} in the limit of high dimensionality is a relation between the mean number $\left < \mathcal{N}_{k} \right >$ of stationary points with instability index $k=0,1,2, \ldots,...$ and the probability distribution of the top $(k+1)$-st eigenvalue in the Gaussian Orthogonal Ensemble  of random matrices. For random energy landscapes \eqref{potential} -- \eqref{covariance} in $N$-dimensions this relation reads
\begin{align}
\label{Nk}
\left < \mathcal{N}_k \right > = \sqrt{\frac{2}{\pi}} \left ( \frac{2}{N} \right )^{N/2} \!\!\! \Gamma \left ( \frac{N+1}{2} \right ) \frac{\sqrt{N}}{m^N} 
\int_{-\infty}^\infty   e^{- N \left[\left (s - \frac{m}{\sqrt{2}} \right )^2 - \frac{s^2}{2}\right]} \rho^{(k+1)}_{N+1} \left ( \sqrt{N} s \right ) ds, 
\end{align}
where  $\rho^{(k+1)}_{N+1}(\lambda) $ is the probability density function of the top $(k+1)$-st eigenvalue in GOE$_{N+1}$, the Gaussian Orthogonal Ensemble of matrices of size $(N+1)\times (N+1)$, and $m$ is the rescaled coupling strength \eqref{m}. To the best of our knowledge,  relation \eqref{Nk} has not been  stated in the literature apart from the case of local minima ($k=0$) \cite{FYODNADAL1}. We derive this relation in  \ref{app0}. Analogous formulae for the fixed energy and the $p$-spin spherical models  expressing the mean number of stationary points with instability index $k$ in terms $\rho^{(k+1)}_{N+1}(\lambda) $ are given in equations \eqref{nkmain} and \eqref{nkmain-pspin}. 

The successful computation of $\left < \mathcal{N}_k \right >$ in the limit of high dimensionality  and that of the mean total number of stationary points $\left < \mathcal{N}_{\text{eq}} \right >=  \sum_{k=0}^N \left < \mathcal{N}_k \right >$  relies on effective approximations of the eigenvalue densities $\rho^{(k+1)}_{N+1}(\lambda)$ and $\rho_{N+1}(\lambda) =\sum_{k=0}^{N} \rho^{(k+1)}_{N+1}(\lambda)$ applicable, depending on the values of $m$, either in the bulk, at the spectral edge, or beyond the spectral edge in the large deviations region. Note that $\rho_{N+1}(\lambda) $ is the usual mean eigenvalue density:  for an infinitesimal $\delta \lambda$, the probability to find an eigenvalue of the GOE$_{N+1}$ matrix in the interval $(\lambda, \lambda+\delta \lambda)$ is given by $\rho_{N+1}(\lambda)  \delta \lambda $. 

Asymptotic analysis in the simplicity region can be performed by making use of the large deviation rate function for  $\rho_{N+1}(\lambda)$, see  \cite{FYODNADAL1,FYOD1}. In this region, asymptotically $\left < \mathcal{N}_0 \right >= \left < \mathcal{N}_{\text{eq}} \right >$ and no approximation of the partial eigenvalue density $\rho^{(k+1)}_{N+1}(\lambda)$ is needed. 

Asymptotic analysis in the hierarchy region 
 can be performed by making use of the Tracy-Widom approximation for eigenvalues at the spectral edge: 
\begin{align}\label{TW}
\rho_{N+1}^{(k+1)}\left (\sqrt{2N} + 2^{-1/2} N^{-1/6} \sigma \right ) \simeq \sqrt{2} N^{1/6} F_k'(\sigma) \, ,
\end{align}
where $F_k(\sigma)$ is the limiting cumulative distribution function of the appropriately scaled top $(k+1)$-st eigenvalue in GOE$_N$ in the limit $N\to\infty$ \cite{TW1,TW2}.  

Asymptotic analysis in the toppling and complexity regions can be performed by making use of the Gaussian approximation for eigenvalues at the spectral edge and in the bulk \cite{GUSTAV1,OROURKE1}. If $k=O(N^{\gamma})$ with $\gamma\in (0,1)$ (spectral edge) then 
\begin{align}\label{normal}
 \rho^{(k+1)}_{N+1}(\sqrt{N} s) \simeq \frac{1}{\sqrt{2\pi \sigma_k^2}} \, \exp \left[-\frac{\big(\sqrt{N} s -\mu_k\big)^2}{2\sigma_k^2}\right]\, ,
 \end{align}
where $\mu_k = \sqrt{2N}\Big[1- \big( \frac{3\pi k}{4\sqrt{2} N}\big)^{2/3}\Big]$ and $\sigma_k^2 = \frac{2\log k }{N^{1/3} (12\pi k)^{2/3}}$. And if  $k=O(N)$ (bulk) then \eqref{normal} holds with the mean value $\mu_k=q_{k}$ and variance $\sigma_k^2=\frac{\log N}{2N(1-q_k^2)}$ expressed in terms of the $k$-th quantile $q_k$ of the Wigner's semicircle law. 

The utility of approximations \eqref{TW} -- \eqref{normal} is in that they allow for an asymptotic evaluation of the integral on the right-hand side in \eqref{Nk}. In this way one obtains the basic building blocks which in turn comprise distinct fractional probabilities for each scaling regime in the random energy landscape model. For details we refer the reader to \ref{app0}. Using a similar approach, the fractional probabilities are obtained for the fixed energy model in \ref{fixeden} and for the $p$-spin spherical model  in \ref{spherical}.

\section{Conclusions}

In this work we propose a detailed picture of the glass--like transition in the random energy landscape model described through the lens of  populations of stationary points. To this end, we work out fractional probabilities of populations of stationary points with fixed number of unstable directions. These fractional probabilities can also be thought of as representing the annealed probability distribution of the instability index of a stationary point picked up at random from the totality of all stationary points. The behaviour of fractional probabilities changes as the system transits from the simple phase with typically very few stationary points to the complex phase with a multitude of stationary points. When exiting the simple phase, the system develops a hierarchy of stability defined as monotonic behaviour of fractional probabilities -- the most probable stationary points are local minima, then stationary points with one unstable direction, and so on. This order of stationary points breaks down as the system enters the complex phase, the process that we refer to as the \emph{toppling of stability hierarchy}. In the vicinity of the transition, we identify toppling  as breaking the monotonicity of fractional probabilities of populations or, equivalently, as  a change in the hierarchy where the local minima cease to be the most probable stationary points.
 
Although our analysis is based mainly on one toy model, we argue that the discussed toppling mechanism is likely to be a universal feature of glass-like transition. To this end, we analyse the constrained random energy landscape model \eqref{potential} with the fixed energy constraint modifying the vicinity of the glassy transition and the spherical spin-glass model \eqref{spinglass} introducing a toppling of dual hierarchy. In both cases we find the same  toppling of stability hierarchy driving transition between the simple and complex phases. We believe this is a generic feature - considered models are all linked to properties of the underlying random matrices which in turn have known universal properties \cite{OXFORD}. At the same time, we would like to emphasize that the transition picture is dependent on two reasonable yet simplifying assumptions of the annealed approximation \eqref{pk} and the zero-temperature limit. Abandoning these assumptions can introduce new phenomena. As an example, although the toppling phenomenon is also expected to happen for the mixed spherical $p$-spin model considered in \cite{FFRT1}, working at finite temperatures introduces new dynamical behaviour. 

Finally, systems with asymmetric couplings have recently attracted considerable interest due to their relevance in various contexts, e.g., neural networks or biological and ecological systems. Our approach relies on the exact relation \eqref{Nk} between the average number of saddles with instability index $k$ and the density of the $(k+1)$-th eigenvalue of the random matrix in the Kac-Rice integral. This allows for a quantitative of analysis of the transition region in the annealed approximation. In the asymmetric setting no such relation is unknown. This makes extending our approach to the asymmetric settings, such as non-gradient random flows (cf \eqref{b1}) 
\begin{align}\label{b1asym}
\dot{\mathbf{x}} = - \mu \mathbf{x} +\mathbf{f} (\mathbf{x}),
\end{align}
challenging. Away from the transition region, one can use tools of the large deviation theory for non-Hermitian matrices and work out the annealed density of the instability index  for the random flow \eqref{b1asym} \cite {KFBA2}. Interestingly, in the left tail of the transition region this density is given by exactly the same functional expression as one found in this paper in the toppling region. Therefore, one may expect the toppling mechanism at work for the random flows \eqref{b1asym} and their variations \cite{FFI,LF}. However, it is unclear whether the intuition based on our annealed calculations in the symmetric case could be helpful in the asymmetric world at large.
 For example, recent work \cite{ARCB2021} on the generalized Lotka-Volterra model with random symmetric interactions argues that adding weak asymmetric interactions completely wipes out marginally stable states and replaces locally stable states with chaotic attractors.  

\ack
We thank F. Bornemann for sharing with us the package \textit{RMTFredholmToolbox} \cite{BORNEMANN} used to calculate Tracy-Widom distributions.

\appendix

\section{ Random energy landscape model}
\label{app0}
In this appendix  we provide details of our calculations in the case of the random landscape model \eqref{potential}. 

\subsection{Calculating $\left < \mathcal{N}_k \right >$ or the mean number of stationary points with index $k$}
\label{derivNk}
Counting statistics for populations of stationary points are given in terms of a formal density 
\begin{align}
\rho_k(\textbf{x}) = \sum_{\textbf{x}^{(k)}_*} \delta (\textbf{x} - \textbf{x}^{(k)}_* )\, ,
\label{rho}
\end{align}
where the summation is over all  stationary points $\textbf{x}^{(k)}_*$ with fixed instability index $k$. This density is in turn expressed by the celebrated Kac-Rice formula 
\[
\rho_k(\textbf{x}) = \left| \det \partial_{ij} E(\textbf{x}) \right| \Theta_k \left( \partial_{ij} E(\textbf{x}) \right) \delta \left( \partial_{i} E(\textbf{x}) \right),
\]
where Heaviside function $\Theta_k$ is equal to $1$ when the Hessian $\partial_{ij} E(\textbf{x})$ has exactly $k$ negative eigenvalues and $0$ otherwise. We defined three basic counting statistics:
\begin{itemize}
\item the total number of stationary points, $\mathcal{N}_{\text{eq}} = \sum_{k=0}^N \int \rho_k(\textbf{x}) d\textbf{x}$ ;
\item the number of stationary 
points with instability index $k$, $\mathcal{N}_k = \int \rho_k(\textbf{x}) d\textbf{x}$;
\item the number of stationary 
points with instability index up to $k$, $\mathcal{N}^{(k)} = \sum_{n=0}^{k} \int \rho_n(\textbf{x}) d\textbf{x}$.
\end{itemize}

Firstly, we combine the density of stationary points $\rho_k(\textbf{x})$ introduced in  \eqref{rho} with the multidimensional Kac-Rice formula:
\begin{align*}
\left < \mathcal{N}_k \right > = \int d\textbf{x} \Big < \left | \det \partial_{ij} E(\textbf{x}) \right | \Theta_k \left ( \partial_{ij} E(\textbf{x}) \right ) \prod_{i=1}^N \delta \left (\partial_{i} E(\textbf{x}) \right ) \Big >_V,
\end{align*}
where $E(\textbf{x}) $ is the random function introduced in  \eqref{potential} and the Heaviside function $\Theta_k$ outputs $1$ when the Hessian has exactly $k$ negative eigenvalues and $0$ otherwise. Averaging is taken wrt. random field $V$.

Following \cite{FYOD1}, the averaging factorize into two terms since the derived fields $\partial_i V$ and $\partial_{ij}V$ decouple as evidenced by the vanishing of their cross-correlations $\left < \partial_i V \partial_{kl} V \right > = 0$:
\begin{eqnarray}
\label{avtotal}
\fl \Big < \left | \det \partial_{ij} E(\textbf{x}) \right | \Theta_k \left ( \partial_{ij} E(\textbf{x}) \right ) \delta \left (\partial_{i} E(\textbf{x}) \right ) \Big >_V = \\
\nonumber 
 \Big < \left | \det \partial_{ij} E(\textbf{x}) \right | \Theta_k \left ( \partial_{ij} E(\textbf{x}) \right ) \Big >_V \Big < \delta \left (\partial_{i} E(\textbf{x}) \right ) \Big >_V.
\end{eqnarray}
The first term is re-expressed using $\partial_{ij} E = \mu \delta_{ij} + \partial_{ij} V$ and an $\textbf{x}$ independent matrix $M$:
\begin{align}
\label{av0}
\Big < \left | \det \partial_{ij} E(\textbf{x}) \right | \Theta_k \left ( \partial_{ij} E(\textbf{x}) \right ) \Big >_V = \Big < \left |\det \left ( \mu - M) \right ) \right | \Theta_k \left ( \mu - M \right ) \left < \delta (\partial V(\textbf{x}) + M ) \right >_V \Big >_M.
\end{align}
where the average $\left < \delta (\partial V(\textbf{x}) + M ) \right >_V$ is computed by first representing a multidimensional delta function in terms of the Fourier integral and then integrating out the $V$ dependent term by using an identity $\left < \left [ \text{Tr} P\partial V \right ]^2 \right > = \frac{\mu_c^2}{N} \left (2\text{Tr} P^2 + (\text{Tr} P)^2 \right )$ with $\mu_c = \sqrt{f''(0)}$. Due to the Gaussianity of the field $V$, the integral does not depend on $\textbf{x}$ and is given by
\begin{align}
\left < \delta (\partial V(\textbf{x}) + M ) \right >_V \sim \exp \left\{ - \frac{N}{4\mu_c^2} \left [ \text{Tr} M^2 - \frac{1}{3} (\text{Tr} M)^2 \right ] \right \}
\end{align}
which is the joint pdf for the random matrix $M$ that we average over in \eqref{av0}. This random matrix is closely related to the GOE since $M$ is real and symmetric. Although the second term $\sim (\text{Tr} M)^2$ is non-standard, the formula can be recast into a proper GOE through introduction of an additional Gaussian integration and a trivial rescaling. The result reads:
\begin{eqnarray}
\label{firstfin}
\fl 
\Big < \left | \det \partial_{ij} E(\textbf{x}) \right | \Theta_k \left ( \partial_{ij} E(\textbf{x}) \right ) \Big >_V =
\\
\nonumber
 \sqrt{\frac{N}{2\pi}} \int_{-\infty}^\infty dt e^{-\frac{N}{2} t^2} \Big < \left |\det \left ( \mu + \mu_c t - M \right ) \right | \Theta_k \left ( \mu + \mu_c t - M \right )  \Big >_{\text{GOE}},
\end{eqnarray} 
where the constant factor $\sqrt{\frac{N}{2\pi}}$ is specified in the $\mu\to \infty$ limit while the average is taken over the GOE with joint pdf given by $P(M) = c_N^{-1} \exp \left ( -\frac{N}{4\mu_c^2} \text{Tr} M^2 \right )$ and $c_N = \mu_c^{N(N+1)/2} 2^{N/2} \left ( \frac{2\pi}{N} \right )^{N(N+1)/4}$. The second term in \eqref{avtotal} is computed trivially by Gaussian integration:
\begin{align}
\label{second}
\int d\textbf{x} \, \Big < \prod_{i=1}^N \delta \left (\partial_{i} E(\textbf{x}) \right ) \Big > = \int \frac{d\textbf{x}}{(i\sqrt{2\pi} f'(0) )^N}\exp \left ( \frac{\mu^2 \textbf{x}^2}{2f'(0)} \right ) = \frac{1}{\mu^N}.
\end{align}
Since the first term \eqref{firstfin} is independent of $\textbf{x}$, we integrate out the space-like variable $\textbf{x}$ as long as $f'(0)<0$.
Lastly, we bring together both terms \eqref{firstfin} and \eqref{second} and find:
\begin{align}
\label{form1}
\left < \mathcal{N}_k \right > = \frac{1}{\mu^N} \sqrt{\frac{N}{2\pi}} \int_{-\infty}^{\infty} dt e^{-\frac{N}{2}t^2} K_{k,N}(z_t),
\end{align} 
where $z_t = \mu +\mu_c t$ and  the matrix-averaged term reads:
\begin{align}
\label{matrixav}
K_{k,N}(z) = \left < |\det ( z-M ) | \Theta_{k}(z-M) \right >_{\text{GOE}},
\end{align}
Formula \eqref{form1} is a generalization to $k\neq 0$ of equation found in \cite{FYODNADAL1}. The Heaviside function is a symmetrized product each for eigenvalues of $M$:
\begin{align*}
\Theta_{k}(M) = \sum_\sigma \theta \left (-\lambda_{\sigma(1)} \right ) \cdots \theta \left (-\lambda_{\sigma(k)} \right ) \theta\left (\lambda_{\sigma(k+1)} \right )\cdots \theta\left (\lambda_{\sigma(N)} \right ).
\end{align*}
It conditions the matrix $M$ to have exactly $N-k$ positive and $k$ negative eigenvalues. To compute $K_{k,N}$ we follow the standard approach of random matrix theory \cite{OXFORD} and change integration variables from matrix elements $M_{ij}$ to its eigenvalues $\lambda_i$:
\begin{eqnarray}
\label{eq1}
\fl 
K_{k,N}(z) = z_N^{-1} \binom{N}{k}  \int d\lambda_1 \cdots \int d\lambda_N \prod_{i<j} |\lambda_i - \lambda_j| \prod_{i=1}^N |z - \lambda_i| e^{- \frac{N}{4\mu_c^2} \lambda_i^2} \times \\
\nonumber  \prod_{i=1}^k \theta\left (-z+\lambda_{i} \right ) \prod_{i=k+1}^N \theta \left (z-\lambda_{i} \right ).
\end{eqnarray}
The binomial is a combinatorial factor resulting from the symmetrization of the Heaviside step function while $z_N = c_N 2^{N} \pi^{-\frac{N(N+1)}{4}} \prod_{i=1}^N \Gamma(1 + i/2) = \left ( 2\sqrt{2} \right )^N \left ( \frac{2\mu_c^2}{N} \right )^{N(N+1)/4} \prod_{j=1}^N \Gamma(1+j/2)$ is the new normalization arising by integrating out the eigenvectors. In  \ref{app1} we derive the formula:
\begin{align}
\label{relation}
K_{k,N}(z) = C_N e^{\frac{Nz^2}{4\mu_c^2}} \rho_{N+1}^{(k+1)}\left (z \sqrt{N/(2\mu_c^2)} \right ) , 
\end{align}
where $\rho_{N+1}^{(k+1)}$ is probability density function of finding the $(k+1)$-th largest eigenvalue of a random matrix of size $(N+1)\times (N+1)$ and $C_N = \sqrt{2} \left ( \frac{2}{N} \right )^{N/2} \mu_c^N \Gamma \left (\frac{N+1}{2} \right ) $. We plug it back to  \eqref{form1} with rescaling $t \to \sqrt{2} t - m$ and obtain the final form given in  \eqref{Nk}:
\begin{align}
\label{Nk0}
\left < \mathcal{N}_k \right > = c_N m^{-N} \sqrt{N} \int_{-\infty}^\infty ds e^{- N f(s;m)} \rho^{(k+1)}_{N+1} \left ( \sqrt{N} s \right ), 
\end{align}
where $\mu/\mu_c = m$, $c_N = \sqrt{\frac{2}{\pi}} \left ( \frac{2}{N} \right )^{N/2} \Gamma \left ( \frac{N+1}{2} \right )$ and $f(s;m) = \left ( s - \frac{m}{\sqrt{2}} \right )^2 - \frac{s^2}{2}$. We readily calculate also formula for the cumulative variant $\left < \mathcal{N}^{(k)} \right >$ as a sum:
\begin{align}
\label{Nkmax0}
\left < \mathcal{N}^{(k)} \right > = c_N m^{-N} \sqrt{N} \int_{-\infty}^\infty ds e^{- N f(s;m)} \sum_{n=0}^k \rho^{(n+1)}_{N+1} \left ( \sqrt{N} s \right ).
\end{align}
We stress that both formulas \eqref{Nk0} and \eqref{Nkmax0} are exact. 

\subsection{Derivation of  \eqref{relation}}
\label{app1}
We establish a relation \eqref{relation}:
\begin{align*}
K_{k,N}(z) = C_N e^{\frac{Nz^2}{4\mu_c^2}} \rho_{N+1}^{(k+1)}\left (z \sqrt{N/(2\mu_c^2)} \right ) , 
\end{align*}
where $C_N =\sqrt{2} \left ( \frac{2}{N} \right )^{N/2} \mu_c^N \Gamma \left (\frac{N+1}{2} \right )$. We start off from the l.h.s. given by \eqref{eq1}:
\begin{eqnarray*}
\fl 
K_{k,N}(z) = z_N^{-1} \binom{N}{k} \int d\lambda_1 \cdots \int d\lambda_N \prod_{i<j} |\lambda_i - \lambda_j| \prod_{i=1}^N |z - \lambda_i| e^{- \frac{N}{4\mu_c^2}
\lambda_i^2} 
 \times \\
 \prod_{i=1}^k \theta\left (-z+\lambda_{i} \right ) \prod_{i=k+1}^N \theta \left (z-\lambda_{i} \right ),
\end{eqnarray*}
where $z_N = \left ( 2\sqrt{2} \right )^N \left ( \frac{2\mu_c^2}{N} \right )^{N(N+1)/4} \prod_{j=1}^N \Gamma(1+j/2)$.
We integrate out the Heaviside functions, rescale $\lambda_{i} =  \sqrt{\frac{2\mu_c^2}{N}} \mu_{i}$ and set $z = y \sqrt{\frac{2\mu_c^2}{N}}$ to find:
\begin{align}
\label{eq4}
K_{k,N}(z) & = z_N^{-1} \left ( \frac{2\mu_c^2}{N} \right )^{N(N-1)/4+N} \tilde{\kappa}_{k,N}(y),
\end{align}
with the rescaled quantity given by:
\begin{eqnarray*}
\fl 
\tilde{\kappa}_{k,N}(y) = (-1)^k  \binom{N}{k} \times \\
\int_y^\infty d\mu_{1} \cdots \int_y^\infty d\mu_{k} \int_{-\infty}^y d\mu_{k+1} \cdots \int_{-\infty}^y d\mu_{N} \prod_{i<j} \left |\mu_{i} - \mu_{j} \right | \prod_{i=1}^N \left ( y - \mu_{i} \right ) e^{- \frac{\mu_i^2}{2} }.
\end{eqnarray*}
This formula is related to the probability that exactly $k$ eigenvalues lie inside an interval $J = (y,+\infty)$ (Definition 8.1 in \cite{FORRESTER}):
\begin{eqnarray}
\label{eq5}
\fl 
E_N(k,J) = \frac{1}{Z_{0,N}(-\infty) }  \binom{N}{k} \times \\
\nonumber 
\int_y^\infty d\mu_{1} \cdots \int_y^\infty d\mu_{k} \int_{-\infty}^y d\mu_{k+1} \cdots \int_{-\infty}^y d\mu_{N} \prod_{i<j} \left |\mu_{i} - \mu_{j} \right |  e^{- \frac{\mu_i^2}{2} } = \frac{Z_{k,N}(y)}{Z_{0,N}(-\infty)} ,
\end{eqnarray}
where $Z_{0,N}(-\infty) = z_N\left (\mu_c = \sqrt{N/2} \right ) = \left ( 2\sqrt{2} \right )^N \prod_{j=1}^N \Gamma(1+j/2)$. Taking a derivative of $E_N$ gives the pdf of the $k$-th largest eigenvalue $\rho_N^{(k)}$:
\begin{align*}
\frac{d}{dy} E_N(k,J) & = \rho_N^{(k+1)}(y) - \rho_N^{(k)}(y), \qquad 1 \leq k \leq N-1, \\
\frac{d}{dy} E_N(0,J) & = \rho_N^{(1)}(y), \\
\frac{d}{dy} E_N(N,J) & = -\rho_N^{(N)}(y). 
\end{align*}
In particular, setting $k=0$ gives the pdf of the largest eigenvalue. These formulas are found by using \eqref{eq5} as a probability distribution. We sum first $k+1$ terms to obtain density $\rho_N^{(k)}$:
\begin{align}
\label{eq44}
\frac{d}{dy} \left [ \sum_{l=0}^k E_{N}(l,J)  \right ] = \rho_N^{(k)}(y).
\end{align}
On the other hand, the  derivative of a single probability function $E_{N+1}$ is related to the quantities $\tilde{\kappa}_{k,N}$ through:
\begin{align*}
\frac{d}{dy} E_{N+1}(k,J) = \frac{N+1}{Z_{0,N+1}(-\infty)} e^{-y^2/2} \left [ \tilde{\kappa}_{k,N}(y) - \tilde{\kappa}_{k-1,N}(y) \right ], \qquad 1 \leq k \leq N,
\end{align*}
along with $\frac{d}{dy} E_{N+1}(0,J) = \frac{N+1}{Z_{0,N+1}(-\infty)} e^{-y^2/2} \tilde{\kappa}_{0,N}(y)$. Due to the telescopic property of the derivative, we sum up $k+1$ terms in order to find a formula for a single $\tilde{\kappa}_{k,N}$:
\begin{align}
\label{eq444}
\frac{d}{dy} \left [ \sum_{l=0}^k E_{N+1}(l,J)  \right ] = \frac{N+1}{Z_{0,N+1}(-\infty) } e^{-y^2/2} \tilde{\kappa}_{k,N}(y).
\end{align}
We combine \eqref{eq4}, \eqref{eq44} and \eqref{eq444} and plug back $y = z \sqrt{\frac{N}{2\mu_c^2}}$ to finally arrive at \eqref{relation}:
\begin{align*}
K_{k,N}(z) = C_N e^{\frac{Nz^2}{4\mu_c^2}} \rho_{N+1}^{(k+1)}\left ( z \sqrt{N/(2\mu_c^2) } \right ),
\end{align*}
where constant prefactor is given by 
\begin{align*}
C_N = z_N^{-1} \left ( \frac{2\mu_c^2}{N} \right )^{N(N-1)/4+N} \frac{Z_{0,N+1}(-\infty) }{N+1} =  \sqrt{2} \left ( \frac{2}{N} \right )^{N/2} \mu_c^N \, \Gamma \left (\frac{N+1}{2} \right ) .
\end{align*}

\subsection{Calculating asymptotic forms of $\left < \mathcal{N}_k \right >$ and $\left < \mathcal{N}^{(k)} \right >$ across the transition}
\label{SecNk0Asymptotic}

In this section we derive the asymptotic approximations of the averages \eqref{Nk0} and \eqref{Nkmax0} in four regions detailed in Section \ref{evolution} and Figure \ref{fig1}. All results are summarized in Table \ref{tab3}. 

\subsubsection{Region a) $m>1, m \in O(1)$ and $k\in O(1)$.}
\label{SecNkRegA}

This case is straightforward as it has only one minimum and does not depend on $m$:
\begin{align}
\label{NkRegA}
\left < \mathcal{N}_k (m) \right > \sim \delta_{k,0}, \qquad m > 1,
\end{align}
as was shown in \cite{FYODNADAL1} by the use of large deviation functions of probability density $\rho_{N+1}^{(k)}$.
\subsubsection{Region b) $m = 1 + \delta/ N^{1/3}$, $\delta>0$ and $k\in O(1)$.}
\label{SecRegB}
We first calculate the asymptotics of prefactors and exponential factor in  \eqref{Nk0}:
\begin{align}
\label{prefRegB}
& c_N (1+\delta/N^{1/3})^{-N} e^{-Nf\left (s;1+\delta/N^{1/3} \right )} \sim 2 e^{-N ( \frac{s^2}{2}-\sqrt{2} s+\frac{1}{2}) - 2\delta N^{2/3} + \sqrt{2} N^{2/3} \delta s - \delta^3/3 } = 2 e^{Ng(s;\delta)}.
\end{align}
We then calculate the integral \eqref{Nk0} through the saddle point method. To this end, we find the saddle from $g'(s;\delta)=0$ as $s_* = \sqrt{2}$ and expand all terms around $s = s_* + \frac{1}{\sqrt{2}} \sigma N^{-2/3}$: 
\begin{align}
\label{NkRegB}
\left < \mathcal{N}_k(m=1+ \delta N^{-1/3}) \right > \sim 2 e^{-\delta^3/3} \int_{-\infty}^\infty d\sigma e^{\delta \sigma} F_k'(\sigma), \qquad \delta > 0,
\end{align}
where $\rho_{N+1}^{(k+1)}\left (\sqrt{2N} + \frac{\sigma}{\sqrt{2} N^{1/6}} \right ) \sim \sqrt{2} N^{1/6} F_k'(\sigma)$ is the family of Tracy-Widom distributions \cite{TW1,TW2} for the $(k+1)$-th largest eigenvalue of the GOE. Cumulative mean \eqref{Nkmax0} is given by a sum of $k+1$ contributions:
\begin{align}
\label{NkmaxRegB}
\left < \mathcal{N}^{(k)}(m=1+ \delta/ N^{1/3}) \right > \sim 2 e^{-\delta^3/3} \int_{-\infty}^\infty d\sigma e^{\delta \sigma} \sum_{n=0}^k F_n'(\sigma), \qquad \delta > 0.
\end{align}

\subsubsection{Region c) $m = 1 + \delta /N^{1/2}$, $\delta \in \mathbb{R}$ and $k = \kappa N^{1/4}$.}
\label{SecRegC}
In this region, we use an approximate result found in \cite{GUSTAV1,OROURKE1}. With $k = \kappa N^\gamma$ and for $\gamma \in (0,1)$, the $(k+1)$-th largest eigenvalue of GOE is asymptotically distributed as:
\begin{align*}
\rho^{(k+1)}_{N+1}(\sqrt{N} s) \sim \frac{1}{\sqrt{2\pi \sigma_k^2}}\exp \left ( - \frac{(\sqrt{N} s - \mu_k)^2}{2\sigma_k^2} \right ),
\end{align*}
with mean $\mu_k = \sqrt{2N} \left ( 1 - \left ( \frac{3\pi k}{4\sqrt{2} N} \right )^{2/3} \right )$ and variance $\sigma_k = \sqrt{\frac{2\log k}{N^{1/3} (12\pi k)^{2/3}}}$. To calculate the cumulative mean \eqref{Nkmax0} we first calculate the prefactor as:
\begin{align}
\label{prefRegC}
c_N (1+\delta N^{-1/2})^{-N} e^{-Nf\left (s;1+\delta N^{-1/2} \right )} \sim 2e^{-N \left ( \frac{s^2}{2}-\sqrt{2} s+\frac{1}{2} \right ) - 2\delta \sqrt{N} + \sqrt{2N} \delta s } = 2e^{N g(s;\delta)}.
\end{align}
From $g'(s;\delta) = 0$ we find the saddle-point at $s_* = \sqrt{2}$ and expand integrand around $s= s_* + \sigma/\sqrt{N}$:
\begin{align}
\label{pref1}
N g(s_* + \sigma/\sqrt{N};\delta) \sim - \sigma^2/2 + \sqrt{2} \delta \sigma.
\end{align}
Now we turn to the asymptotic form of the sum inside the integral:
\begin{align*}
S_{\kappa N^\gamma}(\sqrt{N}s) = \sum_{k=0}^{\kappa N^\gamma} \rho^{(k)}_{N+1}(\sqrt{N} s) 
\end{align*}
in the general case $\gamma \in (0,1)$. Since we inspect the $N\to \infty$ asymptotics, we approximate the sum by the Euler-Maclaurin formula:
\begin{align*}
S_{\kappa N^\gamma}(\sqrt{N}s) \sim N^\gamma \int_0^\kappa d\lambda \rho^{(\lambda N^\gamma)}_{N+1} (\sqrt{N} s) = D_N \int_0^\kappa d\lambda e^{\frac{N^{\frac{2}{3}(2+\gamma)}}{\log N} f(\lambda)} g(\lambda),
\end{align*}
where $D_N = \sqrt{\frac{c_1}{\pi}} \frac{N^{(1 + 8\gamma)/6}}{ \sqrt{\log N}}$, $f(\lambda;s) = - c_1 \left (s- \sqrt{2} + c_2 \lambda^{2/3} N^{2/3(\gamma-1)} \right )^2 \lambda^{2/3}$ and $g(\lambda) = \lambda^{1/3}$ with constants $c_1 = \frac{1}{4\gamma} (12\pi)^{2/3}$ and $c_2 = \sqrt{2} \left ( \frac{3\pi}{4\sqrt{2}} \right )^{2/3}$. In the next step we change variables $\lambda^{2/3} = x$, $\lambda^{1/3} d \lambda = \frac{3}{2} x dx$:
\begin{align*}
S_{\kappa N^\gamma}(\sqrt{N}s) \sim \frac{3}{2} D_N \int_0^{\kappa^{2/3}} xdx e^{\frac{N^{\frac{2}{3}(2+\gamma)}}{\log N} \tilde{f}(x;s)}, 
\end{align*}
where $\tilde{f}(x;s) = f(\lambda=x^{3/2};s)$. Since we will eventually expand the sum around $s = s_* + \sigma/\sqrt{N}$, a natural scale is such that $\tilde{f}(x;\sqrt{2} + \sigma/N^{1/2}) = N^{-1} \tilde{f}_0(x;\sigma)$. This happens when $2/3(\gamma - 1) = -1/2$ or the powers of $N$ in both terms agree. From now on we set $\gamma = 1/4$:
\begin{align*}
S_{\kappa N^{1/4}}(\sqrt{N}s) \sim \frac{3}{2} D_N \int_0^{\kappa^{2/3}} xdx ~e^{\frac{\sqrt{N}}{\log N} \tilde{f}_0(x;\sigma)},
\end{align*}
where $\tilde{f}_0(x;\sigma) = - c_1 \left (\sigma + c_2 x \right )^2 x$. Saddle point in above integral is given by $x_* = - \frac{\sigma}{c_2}$ and must lie within the integration interval $x_* \in (0,\kappa^{2/3})$ otherwise its leading order contribution vanishes. We set $x = x_* + y \left ( \frac{\sqrt{N}}{\log N} \right )^{-1/2}$ and compute the resulting integral:
\begin{align*}
S_{\kappa N^{1/4}}(\sqrt{N}s) \sim \frac{3}{2} D_N e^{\frac{\sqrt{N}}{\log N} \tilde{f}_0(x_*;\sigma)} x_*\sqrt{\frac{2\pi}{-\tilde{f}_0''(x_*;\sigma)}} \frac{\sqrt{\log N}}{N^{1/4}} \theta (x_*) \theta(\kappa^{2/3} - x_*).
\end{align*}
Since $\theta (x_*) = \theta (-\sigma), \theta(\kappa^{2/3} - x_*) = \theta (\sigma + c_2 \kappa^{2/3})$, $\tilde{f}_0(x_*;\sigma) = 0$ and $\tilde{f}_0''(x_*;\sigma) = 2c_1 c_2 \sigma$. Finally, we obtain
\begin{align}
\label{sum0}
S_{\kappa N^{1/4}}(\sqrt{N}s) \sim 2^{3/4} N^{1/4} \frac{1}{\pi} \sqrt{-\sigma} \theta (-\sigma)\theta (\sigma + c_2 \kappa^{2/3}).
\end{align}
We combine \eqref{pref1} and \eqref{sum0} and plug them back to  \eqref{Nkmax0} which result in the final formula given in Table \ref{tab3}:
\begin{align}
\label{NkmaxRegC}
\left < \mathcal{N}^{(\kappa N^{1/4})}(m = 1+\delta/N^{1/2}) \right > \sim 2N^{1/4} \frac{2^{3/4}}{\pi} \int^{c_2 \kappa^{2/3}}_{0} d\sigma \sqrt{\sigma} e^{-\frac{\sigma^2}{2} - \sqrt{2} \sigma \delta},
\end{align}
where $c_2 = \sqrt{2} \left ( \frac{3\pi}{4\sqrt{2}} \right )^{2/3}$. For completeness, we also take its derivative:
\begin{align}
\frac{d}{d\kappa} \left < \mathcal{N}^{(\kappa N^{1/4})}(m = 1+\delta/N^{1/2}) \right > \sim 2N^{1/4} \frac{2^{3/4}}{\pi} \left ( \frac{2}{3} c_2^{3/2} e^{-\frac{c_2^2}{2} \kappa^{4/3} - \frac{\sqrt{2}}{2} c_2 \delta \kappa^{2/3}} \right ).
\end{align}

\subsubsection{Region d) $0<m<1, m \in O(1)$ and $k = \kappa N$.}
\label{SecRegD}
Lastly, we consider the complexity region with an extensive index variable. In this case we use a result found in \cite{GUSTAV1,OROURKE1} valid for $k = \kappa N$:
\begin{align*}
\rho^{(k+1)}_{N+1}(\sqrt{N} s) \sim \frac{1}{\sqrt{2\pi \sigma_k^2}}\exp \left ( - \frac{(\sqrt{N} s - \mu_k)^2}{2\sigma_k^2} \right ),
\end{align*}
with $\mu_k = q_k \sqrt{2N}$ and $\sigma_k = \sqrt{\frac{\log N}{2 N (1-q_k^2)}}$. The quantile parameter $q_k = t^{-1}(k/N)$ is the inverse cdf of the Wigner's semicircle law:
\begin{align}
\label{cdfsemicircle}
t(\lambda) = \frac{2}{\pi} \int_\lambda^1 \sqrt{1-x^2} dx = \frac{1}{\pi} \left ( \arccos \lambda - \lambda \sqrt{1-\lambda^2} \right ).
\end{align}
It is expressed in terms of an inverse incomplete beta function defined through $b_{a,b}(B_{a,b}(z)) = z$ and where $b_{a,b}(z) = \int_0^z u^{a-1} (1-u)^{b-1} du$ and $ t^{-1}(x) = 2 B_{\frac{3}{2},\frac{3}{2}}(1-x) - 1$. We follow essentially the same steps as in the toppling region c), the sum in  \eqref{Nkmax0} is asymptotically given by:
\begin{align*}
S_{\kappa N}(\sqrt{N}s) \sim N \int_0^\kappa d\lambda ~\rho^{(N\lambda)}_{N+1}(\sqrt{N} s) \sim \frac{N}{\sqrt{\pi}} \sqrt{\frac{N}{\log N}} \int_0^\kappa d\lambda e^{\frac{N^2}{\log N} \tilde{f}(\lambda)} g(\lambda),
\end{align*}
with a rescaled quantile function $Q_\lambda= q_{\lambda N}$ we denote $\tilde{f}(\lambda) = - (s-Q_{\lambda} \sqrt{2})^2 (1-Q_\lambda^2)$ and $g(\lambda) = \sqrt{1-Q_\lambda^2}$. We find an approximate value for the integral by the saddle point method. First, there are three saddles:
\begin{align*}
\lambda_*^0 & = 1-B_{\frac{3}{2},\frac{3}{2}}\left (\frac{1}{4}(2-\sqrt{2} s) \right ) , \\
\lambda_*^\pm & = 1-B_{\frac{3}{2},\frac{3}{2}}\left (\frac{1}{16}(8+\sqrt{2} s\pm \sqrt{32+2s^2}) \right ),
\end{align*}
where $\lambda_*^0$ is the extremum. Leading order asymptotics is found when $\lambda_*^0 \in (0,\kappa)$, otherwise the integral is subleading. We expand $\lambda = \lambda_*^0 + x \left ( \frac{N}{\sqrt{\log N}} \right )^{-1} $ and, after integrating out the $x$ variable, find
\begin{align*}
S_{\kappa N}(\sqrt{N} s) \sim \frac{N}{\sqrt{\pi}} \sqrt{\frac{N}{\log N}} \frac{\sqrt{\log N}}{N} \sqrt{\frac{2\pi}{-\tilde{f}''(\lambda_*^0)}} e^{\frac{N^2}{\log N} \tilde{f}(\lambda_*^0)} g(\lambda_*^0) \theta(\lambda_*^0) \theta(\kappa - \lambda_*^0).
\end{align*}
We evaluate some of the terms given above:
\begin{align*}
\theta(\lambda_*^0) & = \theta(\sqrt{2}-s), \\
\theta(\kappa - \lambda_*^0) & = \theta(s - \sqrt{2} Q_\kappa), \\
\tilde{f}(\lambda_*^0) & = 0, \\
g(\lambda_*^0) & = \sqrt{1- s^2/2}, \\
f''(\lambda_*^0) & = - \pi^2 .
\end{align*}
We bring these factors together and the sum $S_{\kappa N}(\sqrt{N} s)$ is equal to
\begin{align}
\label{sum1}
S_{\kappa N}(\sqrt{N} s) \sim \sqrt{N} \frac{1}{\pi} \sqrt{2- s^2} \theta(\sqrt{2}-s) \theta(s - \sqrt{2} Q_\kappa)
\end{align}
which is the Wigner's semicircle law truncated at $s=\sqrt{2}Q_\kappa$. We plug back above formula to  \eqref{Nkmax0}:
\begin{align*}
\left < \mathcal{N}^{(\kappa N)}(m) \right > = c_N m^{-N} N \int_{\sqrt{2} Q_\kappa}^{\sqrt{2}} ds e^{-Nf(s;m)}  \frac{1}{\pi} \sqrt{2- s^2} .
\end{align*}
Lastly, the large $N$ contribution to this integral reads:
\begin{align*}
 \int_{\sqrt{2} Q_\kappa}^{\sqrt{2}} ds e^{-Nf(s;m)}  \frac{1}{\pi} \sqrt{2- s^2} & \sim \frac{2}{\sqrt{\pi N}} \sqrt{1-m^2} e^{\frac{Nm^2}{2}} \theta(1-m) \theta(m-Q_\kappa)
\end{align*}
while the prefactor is $ c_N m^{-N} N \sim 2N e^{-N/2-N \ln m}$. Together, they form the final result given in Table \ref{tab3}:
\begin{align}
\left < \mathcal{N}^{(\kappa N)}(m) \right > \sim 4\sqrt{N/\pi} \sqrt{1-m^2} \theta(1-m) \theta(m-Q_\kappa) e^{N\Sigma_{\text{eq}}(m)} , \qquad m \in (0,1),
\label{NkmaxRegD}
\end{align}
where $\Sigma_{\text{eq}}(m) = \frac{1}{2} (m^2-1) - \ln m$. The corresponding pdf is found by differentiation:
\begin{align*}
\frac{d}{d\kappa} \left < \mathcal{N}^{(\kappa N)}(m) \right > =  4\sqrt{N/\pi} \sqrt{1-m^2} \theta(1-m) \delta(\kappa - t(m)) e^{N\Sigma_{\text{eq}}(m)} , \qquad m \in (0,1),
\end{align*}
where we used $\frac{d}{d\kappa} \theta(m-Q_\kappa) = \delta(\kappa - t(m))$. Closely related formula was found by a different approach in \cite{BRAYDEAN1,FYODWILL1}.

\subsection{Calculating asymptotic forms of $\left < \mathcal{N}_{\text{eq}} \right >$ across the transition}
\label{Nec}

To obtain the mean number of all stationary points we can either follow the same route as in the previous section when deriving the fixed index case or simply use the fact that it is a special case of the cumulative distribution \eqref{Nkmax0}. For $k=N$, we find $\mathcal{N}_{\text{eq}} = \mathcal{N}^{(N)}$ since the sum of the individual eigenvalues pdf's summed over all eigenvalues gives the total density $\sum_{k=0}^N \rho^{(k+1)}_{N+1} = \rho_{N+1}$:
\begin{align}
\label{Neq0}
\left < \mathcal{N}_{\text{eq}} \right > = c_N m^{-N} \sqrt{N} \int_{-\infty}^\infty ds e^{- N f(s;m)} \rho_{N+1} \left ( \sqrt{N} s \right ).
\end{align}
We compute the asymptotics of \eqref{Neq0} in all four regions detailed in Section \ref{evolution} and Figure \ref{fig1}. 
 Due to similarities between  \eqref{Neq0} and \eqref{Nk0}, \eqref{Nkmax0}, in many steps we  will reuse formulas obtained in \ref{SecNk0Asymptotic}.

\subsubsection{Region b) $m = 1 + \delta/ N^{1/3}$, $\delta>0$.}
The calculation of $\left < \mathcal{N}_{\text{eq}} \right >$ in this region was found in \cite{FYOD2}. The prefactor was already found in \eqref{prefRegB}. Likewise, the saddle point method applied to the integral produces the same formula \eqref{NkRegB}. The only difference is the integrand which we find in Prop. 9 of \cite{FFG1}:
\begin{align}
\label{rhoedge}
\rho_{N}\left (y = \sqrt{2N} + \frac{\alpha}{\sqrt{2} N^{1/6}} \right ) \sim \sqrt{2} N^{1/6} \rho_{\text{edge}}(\alpha),
\end{align}
where $ \rho_{\text{edge}}(\alpha) = \left ( \text{Ai}'(\alpha) \right )^2 - \alpha (\text{Ai}(\alpha))^2 + \frac{1}{2} \text{Ai}(\alpha) \left (1 - \int_\alpha^\infty \text{Ai}(t) dt \right )$ is the microscopic spectral density of the GOE near the spectral edge. The final formula given in Table \ref{tab3} reads:
\begin{align}
\label{NeqRegB}
\left < \mathcal{N}_{\text{eq}} (m=1+\delta N^{-1/3})\right > \sim 2 e^{-\delta^3/3} \int_{-\infty}^\infty e^{\alpha \delta} \rho_{\text{edge}}(\alpha) d \alpha, \qquad \delta > 0.
\end{align}

\subsubsection{Region c) $m = 1 + \delta /N^{1/2}$, $\delta \in \mathbb{R}$.}
In this region, the prefactor was already found in  \eqref{prefRegB} while the exponential term is given in  \eqref{pref1}. As in  \ref{SecRegC}, the integral over $s$ is found through the saddle point method around $s = \sqrt{2} + \sigma/N^{1/2}$:
\begin{align*}
\sqrt{N} ds ~ \rho_{N+1} \left (\sqrt{N}s \right ) = d\sigma \rho_{N+1} \left (\sqrt{2N} + \sigma \right ) \sim d\sigma ~ N^{1/4}\frac{2^{3/4}}{\pi} \sqrt{-\sigma} \theta(-\sigma),
\end{align*}
where we use the macroscopic GOE spectral density $\rho_N(x) \sim \pi^{-1} \sqrt{2N - x^2}$. Lastly, we collect these coefficients and the final formula given in Table \ref{tab3} reads:
\begin{align}
\label{NeqRegC}
\left < \mathcal{N}_{\text{eq}} (1 + \delta N^{-1/2})\right > \sim 2N^{1/4} \frac{2^{3/4}}{\pi} \int_0^{\infty} \sqrt{\sigma} e^{- \sigma^2/2 - \sqrt{2} \delta \sigma} d\sigma,
\end{align}
which, after rescaling $\sigma \to |\delta| \sigma$, recreates  (102) of \cite{FYOD2}.

\subsubsection{Region d) $0<m<1, m \in O(1)$.}
Finally, we find the mean number of all stationary points in the complexity region. Firstly, we evaluate the integral  \eqref{Neq0}:
\begin{align*}
\int_{-\infty}^{\infty} ds e^{- N f(s;m)} \rho_{N+1} \left (\sqrt{N}s \right )
\end{align*}
using the saddle point method. Solution to $f'(s;m)=0$ gives the saddle $s_* = \sqrt{2} m$ and the integral is expanded around $s = s_* + \sigma N^{-1/2}$:
\begin{align}
\label{tt}
 \int_{-\infty}^{\infty} ds e^{- N f(s;m)} \rho_{N+1} \left (\sqrt{N}s \right ) = \frac{e^{Nm^2/2}}{\sqrt{N}} \int_{-\infty}^\infty d\sigma e^{-\frac{1}{2} \sigma^2} \rho_{N+1} (\sqrt{2N} m + \sigma).
\end{align} 
The spectral density $\rho_{N}$ is again approximated using the Wigner's semicircle law:
\begin{align}
\label{tt2}
\rho_N(\sqrt{2N} m + \sigma) \sim \frac{\sqrt{2N}}{\pi} \sqrt{1-m^2}.
\end{align}
The formula above does not depend on the parameter $\sigma$ so in  \eqref{tt} we are left with a Gaussian integral $\int_{-\infty}^\infty d\sigma e^{-\frac{1}{2} \sigma^2} = \sqrt{2\pi}$. The asymptotic formula for the prefactor of $\left < \mathcal{N}_{\text{eq}} \right >$ reads:
\begin{align}
\label{tt3}
c_N m^{-N} \sqrt{N} \sim 2 \sqrt{N} e^{-N/2 - N \ln m},
\end{align}
Finally, we collect \eqref{tt}, \eqref{tt2} and \eqref{tt3} to reach the final result given in Table \ref{tab3}:
\begin{align}
\label{Neqfin0}
\left < \mathcal{N}_{\text{eq}} (m)\right > \sim 4 \sqrt{N/\pi} \sqrt{1-m^2} e^{N\Sigma_{\text{eq}}(m)}, \qquad m \in (0,1),
\end{align}
where $\Sigma_{\text{eq}}(m) = \frac{1}{2}(m^2-1) - \ln m$. The exponential part of this formula was calculated in (18) of \cite{FYOD1}.

\section{Random energy landscape model with constraint $E_0 = E(\textbf{x})$}
\label{fixeden}
In this section we consider a variant of the toy model \eqref{potential}:
\begin{align}
\label{potentialfixed}
E(\textbf{x}) = \frac{\mu}{2} \textbf{x}^2 + V(\textbf{x}), \qquad \text{with}~E_0 = E(\textbf{x}),
\end{align}
where we restrict to solutions of a fixed energy $E_0$. In particular, we introduce the means $\left < n_k \right >, \left < n_{\text{eq}} \right >$ and $\left < n^{(k)} \right >$ analogous to the quantities introduced in Section \ref{Cohorts} but with an additional term $\delta \big (E_0 - E(\textbf{x}) \big )$. Since both $\left < n_{\text{eq}} \right >$ and $\left < n^{(k)} \right >$ are easily derived from the cumulative mean $\left < n_{k} \right >$, in what follows we consider only the latter quantity. Firstly, we write down its definition:
\begin{align}
\label{nkdef}
\left < n_k(E_0)\right > = \int d \textbf{x} \Big < \rho_k(\textbf{x}) \delta \big (E_0 - E(\textbf{x}) \big ) \Big >_V,
\end{align}
where the $E_0$ dependence is explicitly stated. The mean number of stationary points with instability index $k$ $\left < \mathcal{N}_k \right >$ is related to \eqref{nkdef} through an integral $\left < \mathcal{N}_k \right > = \int dE_0 \left < n_k(E_0)\right >$. By essentially the same steps as in \ref{derivNk} (also \cite{FYODWILL1}), we arrive at the formula:
\begin{align}
\label{nkmain}
\left < n_k \left (E_0 = N\sqrt{f_0} \epsilon_0 \right ) \right > & = c_N m^{-N} \sqrt{N}  \int_{-\infty}^\infty ds e^{-N f(s;m)} G_{N}(s;m,\epsilon_0)   \rho_{N+1}^{(k+1)} \left ( s\sqrt{N} \right ),
\end{align}
with a function $f(s;m) = \left (s-\frac{m}{\sqrt{2}} \right )^2 - s^2/2$ and a constant $c_N = \sqrt{\frac{2}{\pi}} \left ( \frac{2}{N} \right )^{N/2} \Gamma \left ( \frac{N+1}{2} \right )$. The geometric function $G_N$ reads:
\begin{align*}
G_{N}(s;m,\epsilon_0) = g_N \int_0^\infty dR \frac{1}{R}  e^{-N g(R;m) - N h(s,R;m,\epsilon_0) },
\end{align*}
with a constant $ g_N = \sqrt{\frac{2}{\pi f_0 N}} \frac{q}{\sqrt{q^2-1}}  \left ( \frac{Nm^2}{2}\right )^{N/2} \frac{1}{\Gamma(N/2)}$ and functions $g(R;m) = \frac{m^2}{2} R^2 -\ln R$, $h(s,R;m,\epsilon_0) = \frac{1}{2 (q^2 - 1)} \left [ \sqrt{2} s - m + q \epsilon_0 - \frac{m}{2} R^2 \right ]^2$.

We define parameters $m = \mu/\mu_c$ and $q = \mu_c/\tilde{\mu}_c$ with $\mu_c = \sqrt{f''_0}, \tilde{\mu}_c = -\frac{f'_0}{\sqrt{f_0}} $ and $f$ is the correlation function defining the random field $V(\textbf{x})$ with $f^{(k)}_0 = f^{(k)}(0)$.

The cumulative mean $\left < n_k \right >$ given by \eqref{nkmain} is in a form resembling the corresponding quantity for the unconstrained toy model $\left < \mathcal{N}_k \right >$ given by  \eqref{Nk} as closely as possible. The only additional factor is the geometric function $G_N$ as the only term where the energy $\epsilon_0$ enters into the formula. We can check that indeed $\int dE_0 G_N = 1$ and  \eqref{nkmain} is reduced to the previously studied \eqref{Nk}. Also, we reduce to \eqref{Nk} in the limit $q \to \infty$ where also $G_N \to 1$.

Lastly, we readily find an expression for the number of all stationary points at a fixed energy $E_0$:
\begin{align*}
\left < n_{\text{eq}} \left (E_0 = N\sqrt{f_0} \epsilon_0 \right ) \right > & = c_N m^{-N} \sqrt{N}  \int_{-\infty}^\infty ds e^{-N f(s;m)} G_{N}(s;m,\epsilon_0)   \rho_{N+1} \left ( s\sqrt{N} \right ).
\end{align*}

\subsection{Phase space and calculating the total mean $\left < n_{\text{\text{eq}}} \right >$}
We turn to inspecting the phase space which, in contrast to the toy model \eqref{potential}, is two-dimensional as we vary both the coupling strength $m$ and the energy level $\epsilon_0$. We first study the behaviour of $\left < n_{\text{eq}} \left (E_0 = N\sqrt{f_0} \epsilon_0 \right ) \right >$. Since the prefactor $c_N m^{-N} \sqrt{N} g_N \sim \sqrt{N}$, in what follows we calculate only the integral
\begin{align}
\label{Idef}
I(m,\epsilon_0) & = \int_{-\infty}^\infty ds \int_0^\infty dR \frac{1}{R} e^{-N F(s,R;m,\epsilon_0)} \rho_{N+1} \left ( s\sqrt{N} \right ),
\end{align}
with $F(s,R;m,\epsilon_0) = f(s;m) + g(R;m) + h(s,R;m,\epsilon_0)$. We first inspect the macroscopic phase space based on analysis of \eqref{Idef} as summarized in Figure \ref{fig2a}.

\subsubsection{Macroscopic scale $m \in O(1)$ and $m<m_c$.}
We calculate a leading order contribution via the saddle point method. The relevant saddle points are:
\begin{align}
\label{sps}
s_{\text{sp}} = \frac{\Delta(m,\epsilon_0) + q(mq-\epsilon_0)}{\sqrt{2}(1+q^2)}, \qquad R_{\text{sp}} = \frac{\sqrt{\Delta(m,\epsilon_0) - q(mq-\epsilon_0)}}{\sqrt{m}}, 
\end{align}
where $\Delta(m,\epsilon_0) = \sqrt{2(1+q^2)+ q^2(mq-\epsilon_0)^2}$. The correct saddles were chosen so that both $R_{\text{sp}} >0$ and $s_{\text{sp}} >0$. The latter condition $s_{\text{sp}}>0$ is true for values of $m \in \left ( 0, m_c\right )$ with
\begin{align}
\label{m0def}
m_c = 1 + \frac{1+2q \epsilon_0}{2q^2}.
\end{align}
We expand the integral $I(m,\epsilon_0)$ around $s = s_{\text{sp}} + \sigma/N^{1/2}, R = R_{\text{sp}} + \rho/N^{1/2}$:
\begin{align*}
I(m,\epsilon_0) & \sim \frac{1}{\sqrt{N}} e^{-N F_{\text{sp}}},
\end{align*}
where $F_{\text{sp}} = F(s_{\text{sp}},R_{\text{sp}})$
and we skipped the $N$-independent prefactor to focus on the complexity exponent :
\begin{align}
\label{SigmaEq<0form}
\left < n_{\text{eq}}(E_0 = N\sqrt{f_0} \epsilon_0) \right > \sim \sqrt{N} e^{N\Sigma^<_{\text{eq}}(m;\epsilon_0)}, \qquad m < m_c,
\end{align}
with the complexity exponent given by $\Sigma_{\text{eq}}^<(m,\epsilon_0) = -F(s_{\text{sp}},R_{\text{sp}})$ with function $F$ defined in  \eqref{Idef}. Since $m>0$ is positive, when $m_c$ becomes negative, the leading contribution to $\left < n_{\text{eq}} \right >$ vanishes. From $m_c=0$ we find an energy threshold $(\epsilon_0)_{\text{th}} = - \frac{1+2q^2}{2q}$ below which complexity exponent  is always negative (gray dot in the left plot in Figure \ref{fig2a}).

An implicit equation 
\begin{align}
\label{curvesA}
F_{\text{sp}} = F(s_{\text{sp}}(m,\epsilon_0),R_{\text{sp}}(m,\epsilon_0)) = 0
\end{align}
 defines two curves $\epsilon_0=\epsilon_{\pm} (m)$ \eqref{curves} in  the $(\epsilon_0,m)$-plane (see the left plot in Figure \ref{fig2a}) where the complexity exponent $\Sigma_{\text{eq}}$  changes sign and we move between simple and complex regions. Points on these curves are denoted by $((\epsilon_0)_*,m_*)$. In particular, for value $(\epsilon_0)_{*,\max} = - \frac{1}{2q} $ (dashed line in the left plot in Figure \ref{fig2a}) sign change happens at $m_{*,\max} =1$.

For $\epsilon_0 = (\epsilon_0)_{*,\max} $ and in the large $q$ limit, we recreate the complexity exponent for the unconstrained model in the complexity region \eqref{Neqfin0} $\lim_{q \to \infty} \Sigma^<_{\text{eq}} \left (m, \epsilon_0 =-\frac{1}{2q} \right ) = \frac{m^2-1}{2} - \log m = \Sigma_{\text{eq}}$.

\subsubsection{Macroscopic scale $m \in O(1)$ and $m>m_c$.}
In the $m>m_c$ with $m_c$ defined in \eqref{m0def}, we use a large deviation result for the spectral density:
\begin{align*}
\rho_{N+1} \left ( s\sqrt{N} \right ) \sim e^{-N\phi(s)}, \quad \phi(s) = \frac{1}{2} s\sqrt{s^2-2} - \ln \left ( \frac{s + \sqrt{s^2-2}}{\sqrt{2}} \right ), \qquad s > \sqrt{2},
\end{align*}
and compute the resulting integral $I(m,\epsilon_0)$:
\begin{align*}
I(m,\epsilon_0) \sim \int_{\sqrt{2}}^\infty \int_{-\infty}^\infty \frac{dR}{R} ds e^{- N \mathcal{F}}, \qquad \mathcal{F} = F + \phi,
\end{align*}
where $F$ was defined in  \eqref{Idef}. From now on, we track only the exponential part. From $\partial_s \mathcal{F} = 0, \partial_R \mathcal{F} = 0$ we find the relevant saddles as
\begin{align*}
s'_{\text{sp}} & = \frac{1}{2\sqrt{2}q} \left ( (m q - \epsilon_0)(1 + 2q^2) + (1-2q^2)\sqrt{(m q - \epsilon_0)^2-2} \right ), \\
R'_{\text{sp}} & = \sqrt{\frac{q}{m}} \sqrt{mq - \epsilon_0 - \sqrt{(mq - \epsilon_0)^2-2})},
\end{align*}
where from $s'_{\text{sp}}>\sqrt{2}$ we read off the condition $m>m_c$ with $m_c$ given by  \eqref{m0def}. The exponential contribution to the integral is $I(m,\epsilon_0) \sim e^{-N\mathcal{F}_{\text{sp}}}$ with $\mathcal{F}_{\text{sp}} = F(s'_{\text{sp}},R'_{\text{sp}}) + \phi(s'_{\text{sp}})$. The mean total number is given by
\begin{align}
\label{SigmaEq>0}
\left < n_{\text{eq}}(E_0 = N\sqrt{f_0} \epsilon_0) \right > \sim e^{N\Sigma_{\text{eq}}^>(m;\epsilon_0)}, \qquad m>m_c,
\end{align}
where $\Sigma_{\text{eq}}^>(m;\epsilon_0) = -F(s'_{\text{sp}},R'_{\text{sp}}) - \phi(s'_{\text{sp}})$. Function $\mathcal{F}_{\text{sp}}=0$ vanishes for $\epsilon_0 = - \frac{1}{2q}$ and $m>m_c$. For different values of $\epsilon_0$ we find  $\mathcal{F}_{\text{sp}}>0$ and so the complexity exponent  is negative.


\subsubsection{Microscopic scale in the vicinity of $\left (m_{*,\max},(\epsilon_0)_{*,\max} \right )$.}
We expand the complexity exponents $\Sigma_{\text{eq}}^<$ and  $\Sigma_{\text{eq}}^>$ at the boundary $m=m_c$:
\begin{align*}
\Sigma_{\text{eq}}^<(m) \sim \Sigma_{\text{eq}}^<(m_c) + (m-m_c)\Sigma_{\text{eq}}^<(m_c)' + \frac{1}{2}(m-m_c)^2\Sigma_{\text{eq}}^< (m_c)'' + ..., \\
\Sigma_{\text{eq}}^>(m) \sim \Sigma_{\text{eq}}^>(m_c) + (m-m_c)\Sigma_{\text{eq}}^>(m_c)' + \frac{1}{2}(m-m_c)^2\Sigma_{\text{eq}}^>(m_c)'' + ..., \\
\end{align*}
where from explicit formulas for both complexity exponents we find two first terms in the expansion equal $\Sigma_{\text{eq}}^<(m_c) = \Sigma_{\text{eq}}^>(m_c), \Sigma_{\text{eq}}^<(m_c)' = \Sigma_{\text{eq}}^>(m_c)'$ and the discontinuity happens for the quadratic term  $\Sigma_{\text{eq}}^<(m_c)'' \neq \Sigma_{\text{eq}}^>(m_c)''$. Hence, the proper microscopic scaling is
\begin{align*}
m = m_c + \delta/N^{1/2}, \quad \epsilon_0 = (\epsilon_0)_c + \epsilon/N^{1/2}
\end{align*}
with points related by $m_c = 1 + \frac{1 + 2q (\epsilon_0)_c}{2q^2}$. It contains the critical point $(m,\epsilon_0) = \left (m_{*,\max},(\epsilon_0)_{*,\max} \right ) = \left (1,-\frac{1}{2q} \right )$ which we deal with in what follows. In the right plot of Figure \ref{fig2a} we graph a detailed picture of phase space near this critical point.

We expand the integral \eqref{Idef} with the saddle point method around $\left (m_{*,\max},(\epsilon_0)_{*,\max} \right )$:
\begin{eqnarray*}
\fl
I\left (m=1 + \delta/N^{1/2},\epsilon_0 = 
-\frac{1}{2q} + \epsilon/N^{1/2} \right ) \sim \\
\frac{c'_N}{N}  \int_{-\infty}^\infty d\sigma e^{\frac{2\sqrt{2}q(q\delta - \epsilon)}{2q^2-1}\sigma - \frac{2q^2+3}{2(2q^2-1)}\sigma^2} \rho_{N+1} (\sqrt{2N} + \sigma).
\end{eqnarray*}
with constant $c'_N = \sqrt{2\pi}\sqrt{\frac{q^2-1}{2q^2-1}}  e^{-\frac{(\delta - 2q \epsilon)^2}{4(2q^2-1)}}$. We use the Wigner's semicircle law 
\[
\rho_{N+1} (\sqrt{2N} + \sigma) \sim N^{1/4}\,  \frac{2^{3/4}}{\pi}\,  \sqrt{-\sigma} \, \theta(-\sigma)
\]
 and obtain the mean number of minima near the threshold energy $\epsilon_0 = (\epsilon_0)_{*,\max} +\epsilon/\sqrt{N} $:
\begin{eqnarray}
\label{neqRegC}
\fl 
 \left < n_{\text{eq}}(m=m_{*.\max} + \delta/N^{1/2};E_0 = N \sqrt{f_0} (\epsilon_0)_{*,\max} + \sqrt{Nf_0} \epsilon) \right > \sim 
 \\
 \nonumber N^{1/4} \int_0^\infty d\sigma \sqrt{\sigma} e^{-\frac{
\sigma^2}{2} - \frac{a_1}{\sqrt{2a_2}}\sigma},
\end{eqnarray}
with parameters $a_1 = \frac{2\sqrt{2}q(q\delta - \epsilon)}{2q^2-1}$, $a_2 = \frac{2q^2+3}{2(2q^2-1)}$ and an unspecified $N$-independent prefactor. 
\subsubsection{Microscopic scale in the vicinity of $m=m_c$.}
In the previous section we investigated the vicinity of one point in the phase space marked by a gray square in both plots of Figure \ref{fig2a}. Now we look at the behaviour near the line $m=m_c$:
\[
 m = m_c + \frac{\delta}{N^{1/2}},\quad  \epsilon_0 = (\epsilon_0)_c + \frac{\epsilon}{N^{1/2}}.
 \]
  The result reads 
\begin{eqnarray*}
\fl
\left < n_{\text{eq}}\left (m= m_c + \delta/N^{1/2};E_0 = N \sqrt{f_0} ((\epsilon_0)_c + \epsilon/N^{1/2})  \right ) \right > \sim 
\\
e^{N \Delta_0 + \sqrt{N} \Delta_1 + \Delta_2 } \int_0^\infty d\sigma \sqrt{\sigma} e^{- \frac{\sigma^2}{2} - \frac{a_1}{\sqrt{2a_2}} \sigma},  
\end{eqnarray*}
with 
\begin{align*}
\Delta_0 = &- \frac{1}{2} \log m_c - \frac{1}{2}(1-m_c)(1+q^2(1-m_c)), \\
 \Delta_1 = & \frac{1-m_c}{2m_c} \, (2q m_c \epsilon - \delta),\\
  \Delta_2 = & \frac{4q m_c^2\epsilon(\delta - \epsilon q) - \delta^2\, \big[2q^2(m_c^2-1)+1\big]}{4m_c^2\, (2q^2-1)}\, ,
 \end{align*}
and parameters $a_1,a_2$ defined before. For all values of $m$ besides the critical value $m = m_{*,\max} = 1$, the exponent function $\Delta_0<0$ is negative so the overall mean number of stationary points $\left < n_{\text{eq}} \right >$ is exponentially small in $N$. Hence, along the line $m=m_c$ besides $m= 1$, the fractional probabilities \eqref{pk} loose its interpretation when the counting function in the denominator is vanishing asymptotically with $N$. 

\subsection{Calculating cumulative means $\left < n^{(k)} \right >$}
We calculate the cumulative mean $\left < n^{(k)} \right >$ as a sum of \eqref{nkmain}:
\begin{align*}
\left < n^{(k)}\left (m;E_0 = N\sqrt{f_0} \epsilon_0 \right ) \right > & = c_N m^{-N} \sqrt{N} g_N \int_{-\infty}^\infty ds \int_0^\infty \frac{dR}{R} e^{-N F(s,R;m,\epsilon_0)} S_k(s\sqrt{N}),
\end{align*}
where $S_k(s\sqrt{N}) = \sum_{n=0}^k \rho^{(n+1)}_{N+1} \left ( s\sqrt{N} \right )$. We have already found asymptotic approximations of these sums in the two relevant cases with $k=\kappa N^{1/4}$ for  \eqref{sum0} and $k=\kappa N$ in  \eqref{sum1}.

\subsubsection{Macroscopic scale $m<m_c$ and $k = \kappa N$.}
We use the result of \eqref{sum1} $S_{\kappa N} (\sqrt{N} s) \sim \frac{\sqrt{N}}{\pi} \sqrt{2-s^2} \theta(\sqrt{2} - s) \theta(s - \sqrt{2} Q_\kappa)$ where $Q_\kappa$ is the quantile function defined as the inverse cdf of the Wigner's semicircle law \eqref{cdfsemicircle}. We essentially follow the same steps as in \ref{SecRegC} where the asymptotics of $\mathcal{N}^{(k)}$ were found. We combine it with the same saddle point analysis as in the derivation of $\left < n_{\text{eq}} \right >$ asymptotics:
\begin{align}
\label{NKEQ<0form}
\left < n^{(\kappa N)}(m;E_0 = N\sqrt{f_0}\epsilon_0 ) \right > \sim  \theta_{s_{\text{sp}} \in (\sqrt{2} Q_\kappa,\sqrt{2})} e^{N\Sigma^<_{\text{eq}}(m;\epsilon_0)},
\end{align}
where the saddle point $s_{\text{sp}}$ is given by  \eqref{sps}. 

\subsubsection{Microscopic scale near $\left (m_{*,\max},(\epsilon_0)_{*,\max} \right )$ and $k = \kappa N^{1/4}$.}
We use a derivation of $\left < n_{\text{eq}} \right >$ but with the result \eqref{sum0} for $S_{\kappa N^{1/4}}(\sqrt{N} s) \sim 2^{3/4} N^{1/4} \frac{1}{\pi} \sqrt{-\sigma} \theta(-\sigma) \theta(\sigma + c_2 \kappa^{2/3})$. The formula reads
\begin{eqnarray}
\label{nkRegC}
\fl 
\left < n^{(\kappa N^{1/4})}(m=1 + \delta/N^{1/2};E_0 = -\frac{N\sqrt{f_0}}{2q} + \sqrt{Nf_0} \epsilon) \right > \sim 
\\
\nonumber 
N^{1/4} \int_0^{\sqrt{2a_2} c_2 \kappa^{2/3}} d\sigma \sqrt{\sigma} e^{-\frac{
\sigma^2}{2} - \frac{a_1}{\sqrt{2a_2}}\sigma},
\end{eqnarray}
where the notation is the same as in \eqref{neqRegC}.

\subsection{Calculating probabilities $P_k$}
We can calculate the probabilities $P_k$ provided by both total $\left < n_{\text{eq}} \right >$ and cumulative $\left < n^{(k)} \right >$ means for $m<m_c$ and around $m=m_{*,\max}$. These correspond to regions c) and d) introduced in the paper and describe the toppling and complexity regimes respectively. Since the aim of this Appendix is to focus on the toppling mechanism in the fixed energy model \eqref{potentialfixed}, we altogether skip the simplicity and hierarchy regions a) and b) which are analogous to the unconstrained model. 

\subsubsection{Region c) $m = m_{*,\max} + \delta/N^{1/2}, \epsilon_0 = (\epsilon_0)_{*,\max} + \epsilon/N^{1/2}$ and $k = \kappa N^{1/4}$.}

We use \eqref{pk} for the annealed probabilities and simply calculate the ratio of \eqref{neqRegC} and \eqref{nkRegC}:
\begin{align*}
P_{\kappa N^{1/4}} (\delta,\epsilon;q) \sim \frac{\int_0^{C_2\kappa^{2/3}} d\sigma \sqrt{\sigma} e^{-\frac{
\sigma^2}{2} - \Delta \sigma}}{\int_0^{\infty} d\sigma \sqrt{\sigma} e^{-\frac{
\sigma^2}{2} - \Delta \sigma}}
\end{align*}
where the parameters read
\begin{align*}
    C_2 & = \sqrt{\frac{2q^2+3}{2q^2-1}} \sqrt{2} \left ( \frac{3\pi}{4\sqrt{2}} \right )^{2/3}, \quad \Delta = \frac{2q^2 (\delta - \epsilon/q)}{\sqrt{(2q^2-1)(q^2+2)}}.
\end{align*}
The corresponding probability formula for the unconstrained toy model \eqref{potential} given in Table \ref{tab3} has the same functional form with different parameters $C_2,\Delta$. An easy check ensures that these parameters reduce in the large $q$ limit $\lim_{q \to \infty} C_2 = c_2$ and $\lim_{q \to \infty} \Delta = \sqrt{2} \delta$. The tipping point for the toppling mechanism in this model is when $\Delta$ changes sign or when $\delta = \epsilon/q$. The maximum instability index in this model reads $\kappa'_{\max} = \left (-\frac{\Delta}{C_2} \right )^{3/2} = \left (-\frac{2q^2 (\delta - \epsilon/q)}{c_2 \sqrt{(2+q^2)(2q^2 + 3)}} \right )^{3/2}$. Therefore, the underlying mechanism has the same characteristics as in the unconstrained toy model \eqref{potential} shown in Figure \ref{fig1}.

\subsubsection{Region d) $m<m_c$.}
The ratio of \eqref{SigmaEq<0form} and \eqref{NKEQ<0form} gives the probability in the complexity region:
\begin{align}
\label{Pk2}
P_{\kappa N}(m,\epsilon_0;q) \sim \frac{\theta_{s_{\text{sp}} \in (\sqrt{2} Q_\kappa,\sqrt{2})}}{\theta_{s_{\text{sp}} \in (0,\sqrt{2})}}, 
\end{align}
where 
$s_{\text{sp}} = \frac{\Delta(m,\epsilon_0) + q(mq-\epsilon_0)}{\sqrt{2}(1+q^2)}$ and $\Delta(m,\epsilon_0) = \sqrt{2(1+q^2)+ q^2(mq-\epsilon_0)^2}$. We recast the conditions on the saddle point into that for $m,\epsilon$ and so $s_{\text{sp}} > 0$ means $m>0$, $s_{\text{sp}} < \sqrt{2}$ is translated to $m < m_c$ while $s_{\text{sp}} > \sqrt{2} Q_\kappa$ means $m > m_+$ with $m_+$ known only implicitly. For $m\in (0,m_c)$, the denominator in \eqref{Pk2} is always equal to 1 and moreover we can invert the second inequality $\theta(s_{\text{sp}} > \sqrt{2}Q_\kappa ) = \theta \left ( \kappa - t (s_{\text{sp}}/\sqrt{2}) \right )$ with the Wigner's semicircle law cdf given by  \eqref{cdfsemicircle}:
\begin{align*}
t(s) = \frac{2}{\pi} \int_s^1 \sqrt{1-x^2} dx.
\end{align*}
As a check, we study $q \to \infty$ limit where $\lim_{q \to \infty} s_{\text{sp}} = \sqrt{2} m$ which again reduces to the formula found in Table \ref{tab3}. Finally, the cdf reads $P_{\kappa N}(m,\epsilon_0;q) \sim \theta \left (\kappa - t (s_{\text{sp}}/\sqrt{2}) \right )
$ while its pdf is a delta function:
\begin{align*}
\frac{d}{d\kappa} P_{\kappa N}(m,\epsilon_0;q) \sim \delta \left (\kappa - t (s_{\text{sp}}/\sqrt{2}) \right ).
\end{align*}

\section{$p$-spin spherical model}
\label{spherical}

In this section we consider statistics of stationary points in a variant of $p$-spin spherical model following closely \cite{FYOD2}. The energy function we consider reads:
\begin{align*}
E_{\circ}(\textbf{x}) = \sum_{i_1,...,i_p=1}^{N+1} J_{i_1,i_2,...,i_p} x_{i_1} x_{i_2} ... x_{i_p} + \sum_{i=1}^{N+1} h_i x_i,
\end{align*}
where $\textbf{x}$ is now an $N+1$ dimensional vector constrained to lie on the sphere $\sum_{i=1}^{N+1} x_i^2 = N$ and $p\geq 2$ is an integer. The symmetric coupling matrix $J$ and the random external field $h_i$ are both drawn from a Gaussian distribution with mean and variance given by:
\begin{align*}
\left < J_{i_1,i_2,...,i_p} \right > & = 0, \quad \left < (J_{i_1,i_2,...,i_p})^2 \right > = \frac{J^2}{pN^{p-1}}, \\
\left < h_i \right > & = 0, \quad \left < h_i^2 \right > = \sigma^2. \\
\end{align*}
If we treat the energy itself as a random field, its covariance structure reads:
\begin{align*}
\left < E_{\circ}(\textbf{x}) \right > & = 0, \\
\left < E_{\circ}(\textbf{x}) E_{\circ}(\textbf{x}') \right > & = N \tilde{f} \left ( \frac{\textbf{x} \cdot \textbf{x}'}{N} \right ),
\end{align*}
with the correlation function $\tilde{f}(u) = \frac{J^2}{p} u^p + \sigma^2 u$. We follow \cite{FYOD2} and arrive at a formula for the mean counting function:
\begin{align*}
\left < \mathcal{N}_k \right >_{\circ} = \bar{c}_{J,p,\sigma} \sqrt{\frac{N}{2\pi}} \int dt e^{-\frac{N}{2} t^2} K'_{k,N}(z_t),
\end{align*}
where $\bar{c}_{J,p,\sigma} = \frac{2\sqrt{\pi}}{\Gamma \left ( \frac{N+1}{2} \right )} \left ( \frac{2}{N} (J^2 + \sigma^2) \right )^{-N/2} $ and $z_t = t\sqrt{J^2p + \sigma^2}$. Average $K'$ is given by
\begin{align*}
K'_{k,N}(z) = \left < |\det (z - H)| \Theta_k(z-H) \right >_H,
\end{align*}
where $H$ is a matrix drawn from the GOE with jPDF $P(H) \sim \exp \left ( - \frac{N}{4J^2(p-1)} \text{Tr}H^2 \right )$. Lastly, we use a relation \eqref{relation} with replacement $\mu_c^2 \to J^2(p-1)$ stemming from a different normalization of the random matrix and find:
\begin{align*}
K'_{k,N}(z) = C_N e^{\frac{Nz^2}{4J^2(p-1)}} \rho^{(k+1)}_{N+1} \left ( \frac{z}{\sqrt{2J^2(1-p)}} \sqrt{N} \right ),
\end{align*}
with $C_N = \sqrt{2}\left ( \frac{2J^2(p-1)}{N} \right )^{N/2} \Gamma \left (\frac{N+1}{2} \right )$. We plug it back, rescale $t = s \sqrt{\frac{2J^2(p-1)}{J^2p + \sigma^2}}$ and find
\begin{align*}
\left < \mathcal{N}_k \right >_{\circ} = c'_N \int ds e^{ - \frac{N}{2} \frac{J^2(p-2) - \sigma^2}{J^2 p + \sigma^2} s^2} \rho_{N+1}^{(k+1)} \left ( \sqrt{N} s\right ),
\end{align*}
with $c'_N = 2 \sqrt{2N} \sqrt{\frac{J^2(p-1)}{J^2p + \sigma^2}} \left ( \frac{J^2(p-1)}{J^2 + \sigma^2}\right )^{N/2}$. Lastly we introduce a single parameter $B = \frac{J^2(p-2) - \sigma^2}{J^2 p + \sigma^2}$ so that 
\begin{align}
\label{nkmain-pspin}
\left < \mathcal{N}_k \right >_{\circ} & = c'_N \int ds e^{ - \frac{NB}{2} s^2} \rho_{N+1}^{(k+1)} \left ( \sqrt{N} s \right ),
\end{align}
where the constant reads $c'_N = 2 \sqrt{N} \left ( \frac{1+B}{1-B} \right )^{\frac{N+1}{2}} \sqrt{1-B}$. We observe that for $p \geq 2$, the parameter $B \in \left (-1,\frac{p-2}{p} \right ]$. For completeness we also write down the total and the cumulative number of stationary points:
\begin{align*}
\left < \mathcal{N}^{(k)} \right >_{\circ} & = c'_N \int ds e^{ - \frac{NB}{2} s^2} \sum_{n=0}^k \rho_{N+1}^{(n+1)} \left ( \sqrt{N} s \right ), \\
\left < \mathcal{N}_{\text{eq}} \right >_{\circ} & = c'_N \int ds e^{ - \frac{NB}{2} s^2} \rho_{N+1} \left ( \sqrt{N} s \right ).
\end{align*}
Formula for $\left < \mathcal{N}_{\text{eq}} \right >_{\circ}$ agrees with (41) of \cite{FYOD2}.
\subsection{Regions}
According to \cite{FYOD2}, there exist four regions as in the toy model \eqref{potential}. Parameter $B$ combining $J$, $\sigma$ and $p$ serves a role analogous to the coupling strength $m$. The scaling parameters for $B$ and the instability index $k$ in all the regions are summarized in Table \ref{tab7}. As before, regions a) and d) are macroscopic with $B \in O(1)$ and region b), c) are microscopic with $B \in O(N^{-1/3})$ and $B \in O(N^{-1})$. Toppling region has a different scaling in comparison to the toy model \eqref{potential} where $m$ scaled as $O(N^{-1/2})$. Consequently, the instability index scaling is $k = \kappa N$ in regions c) and d). 

\subsection{Results on the total mean number $\left < \mathcal{N}_{\text{eq}} \right >_{\circ}$}
$\left < \mathcal{N}_{\text{eq}} \right >$ across all four regions of change was calculated in \cite{FYOD2}. These results are summarizes in the last column of Table \ref{tab7}. In comparison to Table \ref{tab3}, regions a) and b) look similar although, due to topological constraints, the minimal number of total stationary points is $2$. Regions c) and d) behave similarly to analogous regions in the toy model although the scaling with $N$ in c) is different.

\subsection{Results on the cumulative mean number $\left < \mathcal{N}^{(k)} \right >_{\circ}$}
Although previously we mainly cited \cite{FYOD2}, in this section most formulas are completely or partially new. We expand cumulative means around points of total stability $k=0$ and total instability $k=N$ due to topological constraints present in the system. In particular, this constraint is most clearly manifested in the total number of all stationary points $\left < \mathcal{N}_{\text{eq}} \right >_{\circ}$ being equal to $2$ in the simple region a). Closer inspection conducted in \cite{FYOD2} reveals that stationary points in the simple region are necessarily a total minimum with instability index $k=0$ and a total maximum with $k=N$. This fact has consequences also in the toppling region c). 

\subsection{Calculating probabilities $P_k$}
With the knowledge of $\left < \mathcal{N}^{(k)} \right >_{\circ}$ and $\left < \mathcal{N}_{\text{eq}} \right >_{\circ}$ the annealed probabilities,  \eqref{pk} ,  are easily computed from the results in the previous two sections.  The results are summarized in Table \ref{tab7}. Aforementioned topological constraint is manifested across all regions as symmetries of pdfs upon substitutions $k \to N-k$ and $\kappa \to 1-\kappa$. All results in this section are presented in Figure \ref{fig3} where the symmetry is evident. 

\bigskip

\bibliographystyle{unsrt.bst}
\bibliography{GK}

\end{document}